\documentclass[10pt]{IEEEtran}
\usepackage{amsmath,amsthm,amssymb}
\usepackage{remark}
\usepackage{boxedminipage}
\usepackage{pgfplotstable}
\usepackage{float}
\usepackage{url}
\def\be{\begin{eqnarray}}
\def\ee{\end{eqnarray}}
\def\bee{\begin{eqnarray*}}
\def\eee{\end{eqnarray*}}

\def\la{\langle}
\def\ra{\rangle}
\def\E{\mathbb{E}}
\def\P{\mathbb{P}}

\def\bmx{\begin{pmatrix}}
\def\emx{\end{pmatrix}}

\begin{document}

\newremark{remark}{Remark}

\parindent0mm
\newtheorem{theorem}{Theorem}
\newtheorem{lemma}{Lemma}
\newtheorem{ex}{Example}
\newtheorem{cor}{Corollary}
\newtheorem{defn}{Definition}
\newcommand{\ignore}[1]{}

\title{Distributed Ledger Technology for Smart Cities, The Sharing Economy, and Social Compliance}

\author{P.~Ferraro,\thanks{Pietro Ferraro and Robert Shorten are with the School of Electrical and Electronic Engineering of the
    University College Dublin, Belfield Dublin 4 D04 V1W8, Ireland.}
  C.~King,\thanks{Christopher King is with the Department of Mathematics, Northeastern University, Boston, MA 02115 USA.} and
  R.~Shorten\thanks{A first version of this manuscript is posted to the \emph{Arxiv},  arXiv:1807.00649}}

\maketitle

\begin{abstract}
This paper describes how Distributed Ledger Technologies can be used to enforce social contracts and to orchestrate the behaviour of agents trying to access a shared resource. The first part of the paper analyses the advantages and disadvantages of using Distributed Ledger Technologies architectures to implement certain control systems in an Internet of Things (IoT) setting, and then focuses on a specific type of DLT based on a Directed Acyclic Graph. In this setting we propose a set of delay differential equations to describe the dynamical behaviour of  the Tangle, an IoT-inspired Directed Acyclic Graph designed for the cryptocurrency IOTA. The second part proposes an application of Distributed Ledger Technologies as a mechanism for \emph{dynamic deposit pricing}, wherein the deposit of digital currency is used to orchestrate access to a network of shared resources. The pricing signal is used as a mechanism to enforce the desired level of compliance according to a predetermined set of rules.  After presenting an illustrative example, we analyze the control system and provide sufficient conditions for the stability of the network.

\end{abstract}

\section{Introductory remarks}
\label{Sec: Introduction}
{\em Bitcoin}, and the technology that underpins it, {\em Blockchain}, have recently become a source of great debate and 
controversy in both business and scientific communities.  To its supporters, {\em Distributed Ledger Technology} (DLT) (the agnostic 
term for Blockchain and related technologies)\cite{Walport}\cite{Olnes} is a key technology that will unlock new disruptive business models such as peer to peer trading, protect the rights of individuals, 
democratise society, and remove the need for central arbiters in many applications (the greedy middle that manages and exploits our assets and identities for financial reward). To its detractors, DLT is nothing more than pure hype, irrational speculation, and a means to enable new forms of illegality built on the anonymity that underpins the technology. DLT as a technology is truly unparalleled in its ability to split and divide opinions. Countries such as Switzerland and Singapore are openly embracing its potential, while other countries, such as China and India are trying  to regulate its use \cite{SpringerChina}. Leading societal thinkers are also split on DLT; with George Soros\protect\footnote{https://www.forbes.com/sites/gurufocus/2018/01/25/george-soros-from-davos-bitcoin-is-a-typical-bubble/\#7a1d097929d0}  thinking it to be nothing more than a  bubble and others such as Al Gore embracing the idea that 
algorithms might one day assume some of the functions of government\protect\footnote{https://www.forbes.com/sites/ksamani/2017/10/04/how-crypto-will-reshape-capitalism-as-we-know-it/2/\#5b9daa0e6ed4}.\newline 

Though the schism in DLT thinking is very real, everyone seems to agree on one basic fact - that DLT is potentially a very disruptive technology. Even opponents of the technology are exploring the many ways it can be used, and the consequent potential implications for society. Roughly speaking, as the name suggests,  DLT is a technology for keeping multiple distributed copies of a single ledger. Multiple distributed holders of this ledger achieve consensus to agree on the contents of a ledger, and manage it in a manner so that it cannot be altered. This ledger can be used not only to keep track of financial transactions, but also to record ``{\em who did what, and when}" for a whole host of 
non-financial applications (such as keeping track of food as it passes through a supply chain). The immutable nature of the technology means that DLT is suitable for a vast array of applications in which accurate and honest records of transactions are important. Given the ubiquity of such applications, it is unsurprising that large corporations such as IBM\protect\footnote{https://www.ibm.com/blockchain/what-is-blockchain} and Facebook\protect\footnote{https://www.cnbc.com/2018/05/09/zuckerberg-invests-in-blockchain-to-keep-facebook-relevant.html} are investing heavily in this technology.\newline 

While current applications of DLT are mainly focussed on {\em payments} and on {\em record keeping}, our interest in the technology stems 
from the need to enforce compliance in the sharing economy applications. Motivated by a large class of such systems, in which humans and machines,  or machines and other machines, must orchestrate their behaviour to achieve a common goal, we are interested in using DLT as a design tool. Even though, formally speaking, this use of DLT appears to be new, we believe it to be very significant, and as we shall explain, distributed ledger technologies give rise to a number of properties that make them suitable to solve a problems that arise in the context of {\em Smart Cities}\protect\footnote{{See, for example, the April 2018 issue of {\em Proceedings of IEEE} (special issue on Smart Cities).}}.\newline
\subsection{Motivational Examples: Smart Cities, the Sharing Economy, and Social Contracts}
\label{Subsec: Cyber-phisical}
We are interested in designing a certain type of cyber-physical system. In the following examples, humans and machines must orchestrate 
their behaviour, sometimes sequentially, in order to achieve a common goal - such as sharing of a resource. Even though these examples appear mundane, they represent an important class of common problems that are arising more frequently in the arena of smart cities.\newline

{\em (i) Electric vehicle (EV) charge point anxiety :} Many issues impeding the adoption of electric vehicles, such as long charging times, and range anxiety, have been addressed by advances in technology. One issue, that is more related to human behaviour than technology, remains however.  That is the issue of {\em Charge Point Anxiety}. More specifically, 
public charge points are often occupied in many urban centres either by EV owners parking there for the entire workday (despite the EV being fully charged), or by ICE vehicles illegally occupying designated spaces. In either scenario the charge point is unavailable to other users  resulting in under-utilization of valuable infrastructure. This leads to EV car owners experiencing  the fear that a charge point may not be accessible when needed - {\em Charge Point Anxiety}. This is one of the main remaining barriers preventing the mass uptake of EVs. It is important to note that the inability to connect vehicles to the network is not just an inconvenience for EV owners. It also reduces the ability of the electrical grid to store energy in the EV fleet. This is an important consideration in the design of grid ancillary services - especially for {\em vehicle to grid} (V2G) services and in using the vehicle fleet as a storage buffer.\\

The most intuitive solution to charge point anxiety would be increase the number of charge points in these areas but, unfortunately, this is often not feasible due to cost. An alternative idea is to develop an adapter to extend the reach of charge points and so allow multiple EVs to connect simultaneously. To this end we designed an adapter (called a dockChain Adapter \cite{EVAnxiety}) that allows the adapters to connect to a charge point in a `daisy-chained' or `cascaded' manner as shown below in Figure 1.  It is envisaged that each car will carry an adapter as a standard component, similar to a spare wheel. When connecting to a charge point, the car owner connects the adapter to the charge point (or to another dockChain adapter), then connects the car to his/her adapter  - as depicted in Figure 1.

\begin{figure}[H]
        \begin{center}
                {\includegraphics[width=3.5in]{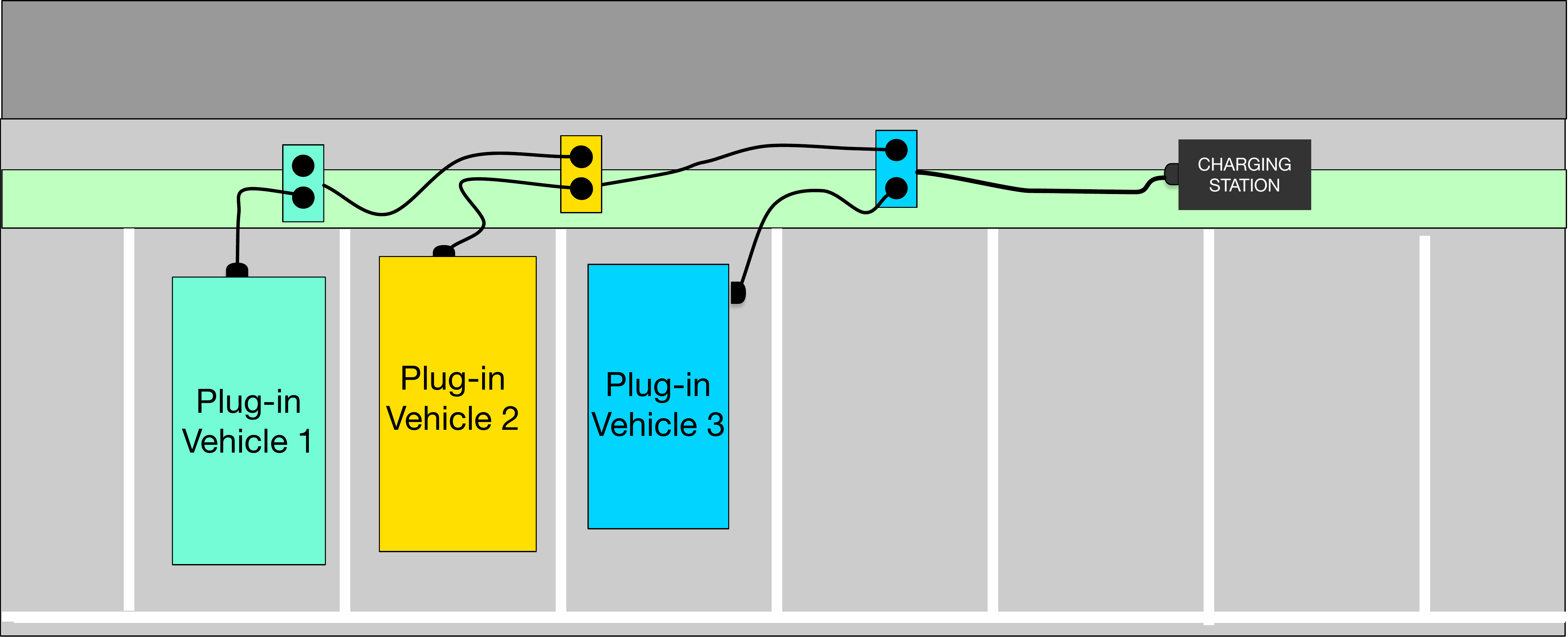}}
                \label{fig:chain}
                \caption{Three vehicles connected to a single charge point and charging simultaneously using three dockChain adapters}
        \end{center}
\end{figure}

To see how DLT technology arises in the context of this example, suppose now that the yellow car wishes to disconnect from the chain. The car owner simply disconnects the car from the yellow box, and then the yellow box from the blue box that is upstream. One must now hope that the owner of the yellow car will connect the green and blue boxes together (assuming that the cables reach). There is, of course, no guarantee  that this happens.  One could use a ratings based system (as used by Uber or Tripadvisor) to rate compliance. An alternative to this approach is to incentivise good behaviour using a crypto-token. That is, to use a digital coin, as part of a deposit system. When the yellow car wishes to remove itself from the chain, it deposits a digital token into the charge point; once it reconnects the chain, its token is returned. In this case the token is used as a bond; if it is of sufficient value, then the yellow car owner is greatly incentivised to reconnect the chain\protect\footnote{{This assumes that cables are of sufficient length to enable compliance. In this context, note that a simple GPS chip in each box can be used to detect if it is possible to reconnect boxes once a car is removed from the chain. The logic for returning a token can be then refined accordingly.  }}.  \newline 


{\em (ii) Smart charging hubs for electric bikes :} While EV's are currently experiencing massive growth in popularity, mainly due to increased awareness around air-quality, they will not solve all mobility problems in cities. First, in many cities, the impact of such vehicles is limited, due to the way in which people live. For example, in cities such as Berlin, Germany, where people mostly reside in apartments, lack of charging infrastructure (or access to it), is likely to impede adoption. More generally, exchanging traditional vehicles for electric vehicles, will not address problems associated with road or parking congestion. It is in this context that electric bikes, or {\em pedelecs}, are seen as an essential component on the path to e-mobility. e-bikes are easily stored, do not require any infrastructure (the battery disconnects from the bike and can be charged from a regular wall socket), and their range (circa 40km) makes them eminently suitable for use in urban environments.  In addition, the on-demand electrical assist provided by the motor, effectively removes many of the usual impediments to cycling (topology, wind, age of cyclist). Furthermore, the opportunity to develop services, for and from such bikes is very appealing, and it is in this context that we are devloping, jointly with MOIXA\protect\footnote{http://www.moixa.com/}, a smart battery unit. Roughly speaking, our smart battery is a unit that aggregates the batteries from a number of e-bikes. For example, this could be part of a mailbox system in an apartment block. Typically, each e-bike battery is of the order of 500 $Wh$; thus the aggregated amount of storage in say 30 batteries would be enough to power ancillary services in an entire building, making such aggregations an attractive part of any backup storage system. Similar opportunities exist for on-street systems. \newline
\begin{figure}[H]
        \begin{center}
                {\includegraphics[width=2.5in]{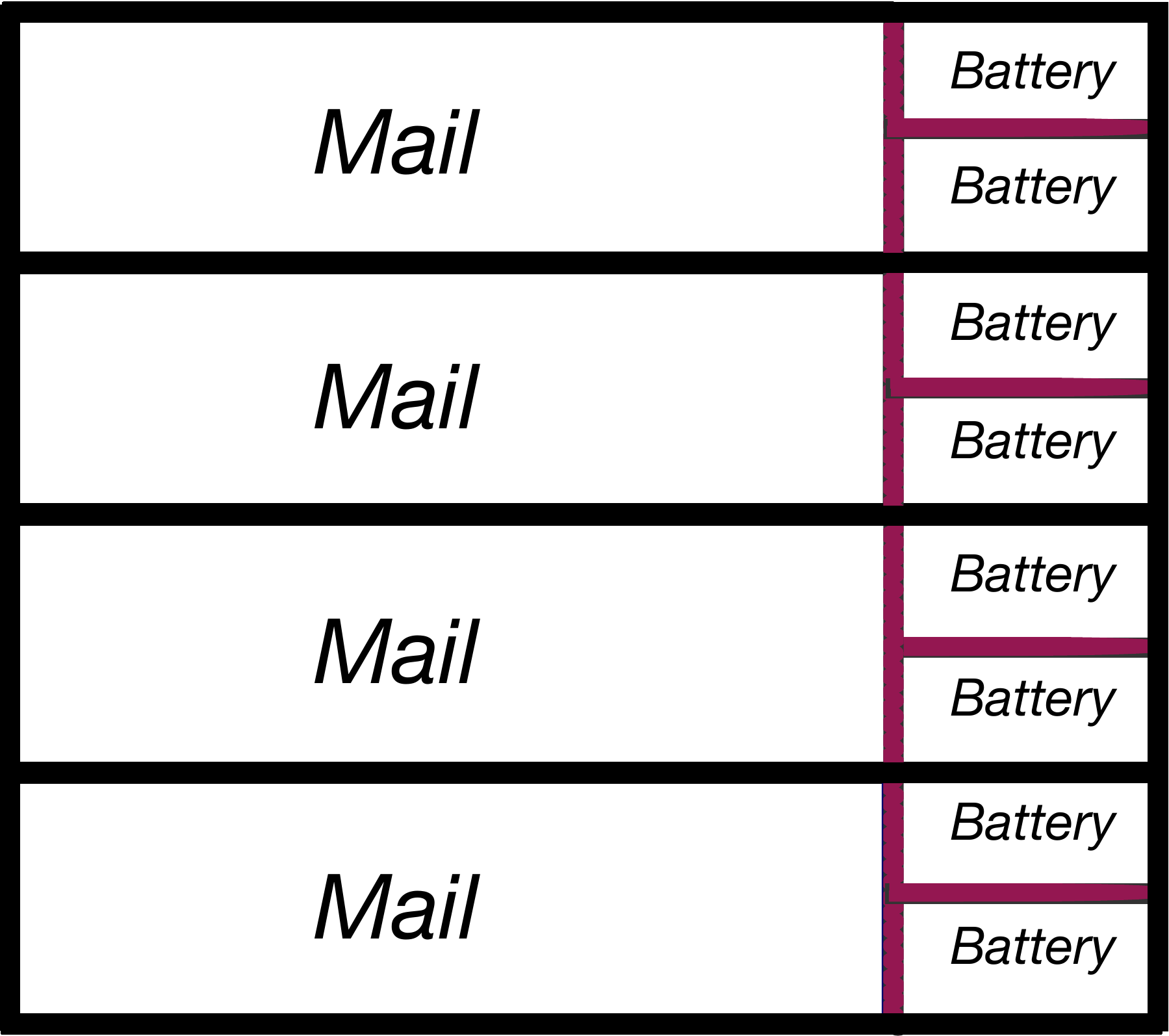}}
                \label{nonproper}
                \caption{A schematic of the ChargeWall system}
        \end{center}
\end{figure}
To see how DLT technology arises in the context of this example, we must first note that the batteries are financially valuable parts of the e-bike system. In our system, as part of a resident management plan, apartment owners would give residents access to an e-bike, as well as access to shared vehicles,  in exchange for fewer parking spaces. Residents would purchase crypto-tokens (whose value would exceed that of a battery) and use a digital deposit system as described above; namely in order to release a battery, users would deposit a token into the {\em ChargeWall}, which would then be returned when the discharged battery is returned.\newline

{\em (iii) The scourge of disposable cups :} It should be clear to the reader that the above problems share certain common features; namely humans, in their interaction with machines, must comply with certain behaviour to achieve common societal goals. Another example of an important system of this type concerns irresponsible waste disposal. For example, in 2011 it was estimated that 2.5 billion coffee cups were thrown away (the figure is likely to be higher now)\protect\footnote{https://www.bbc.com/news/science-environment-43739043}. To many, decomposable cups  are seen as the solution to this problem. Yet, at closer inspection, the introduction of biodegradable cups is only a small part of the solution; namely, not only must we use such cups when drinking our coffee, we must also put our biodegradable cup in the correct trash can or else risk contaminating other waste. As in the previous examples, that is where DLT comes in. One could associate a digital token with an RFID enabled biodegradable (or reusable) cup, and return this token once the user has placed the cup in the correct trash can.  \newline

{\em (iv) General compliance problems :} It should be clear to the reader that there are many smart-city compliance problems where digital deposits can be used to enforce both human and machine behaviour. Examples include enforcing time-restricted parking, interfacing with 3D printers, gaining access to charge points, and ensuring that cyclists and car drivers comply with traffic light signalling (we shall see this last example in much more detail later). {More generally, techniques to enforce compliance with social contracts are a fundamental part of any sharing economy system. The design of shared systems is both topical from a practical and theoretical perspective \cite{sharing1,sharing2}.  Analytic issues associated with developing parctical sharing systems include the following basic questions.\newline 
\begin{itemize}
\item[(i)] How many shared objects are needed to provide population with a certain (acceptable) quality of service \cite{sharing2}?\newline
\item[(ii)] How does one manage access so that everyone gets a fair share of the available resource \cite{sharing3}?\newline
\item[(iii)] How does one regulate during periods of excess demand (and potentially design backup buffers) \cite{sharing1}?\newline
\item[(iv)] How do we enforce compliance so that people behave in a responsible manner? For example, how do we ensure resources are released when not in use, and made available for others. For example, access to charge points, and parking spaces in cities, are examples where time-limited access is often abused.\newline  
\end{itemize}
While items (i)-(iii) have attracted attention in the literature \cite{sharing1,sharing2,sharing3,sharing4}, the very important issue of social compliance has been discussed to a much lesser degree, and it is in this context that we see DLT as a particularly attractive technology.  Assuming that the reader is convinced of the merits of these applications,  the reader may also ask why DLT is the method of choice for implementation. Could one not use any other payment method - such as PayPal or Visa to realise such a system? The answer is that DLT is much more suited to such applications than other digital payment methods for a number of reasons.}\newline 
\begin{itemize}
\item[(a)] First, DLT is a distributed technology making it much more robust than a centralised payment system.\newline 
\item[(b)] Second, at least some DLT technologies (e.g. Legicash,\protect\footnote{https://legi.cash/} Byteball,\protect\footnote{https://byteball.org/} IOTA \cite{Popov}) are designed with the objective to be more suited to high frequency micro-trading than say PayPal or Visa. For example, sometimes, low value transactions will not be processed by traditional digital payment systems due to a lack of incentive in the transaction value for vendors (note: this is also a problem for some DLT architectures such as Bitcoin).\newline 
\item[(c)] Third, PayPal or Visa will always take a transaction fee - making their use in digital deposit based systems questionable; namely, where the entire value of the token is intended to be returned to a compliant agent.\newline
 \item[(d)] Finally, in principle, DLT tokens are more like cash than other digital forms of payment. More specifically, transactions are pseudo-anonymous\protect\footnote{https://laurencetennant.com/papers/anonymity-iota.pdf}, that is the cryptographic nature of the addressing is less revealing that other forms of digital payments that are uniquely associated with an individual, the time and location of the spend, and the item that was transacted. This is unlike card based transactions, which always leave a trail of what was done and when. Thus from a privacy perspective (re. Cambridge Analytica and Facebook\protect\footnote{https://www.bbc.com/news/topics/c81zyn0888lt/facebook-cambridge-analytica-data-scandal}), the use of DLT is much more satisfactory than traditional digital transactions.\newline
\end{itemize}  
From the perspective of social compliance, {\bf fee-less transactions} and {\bf privacy} are the key considerations. 

\subsection{Related work}

Since the publication of the white paper where Blockchain was first introduced \cite{Nakamoto}, the literature on DLTs  has grown rapidly as researchers from industry and academia alike have been trying to explore the limits of this new technology and its possible applications.  Nowadays it is possible to find plenty of material to understand the underlying principles on which DLTs are built: relevant reviews on the functioning principles of the Blockchain have been presented in \cite{Puthal}, \cite{Conoscenti} and \cite{Zheng}, while a thorough exposition on a more recent architecture for DLTs, called the Tangle, can be found in \cite{Popov} and \cite{Tangle}. In \cite{Gatteschi} the authors discuss the advantages and disadvantages of Blockchain taking the insurance sector as an example. Unlike much of the literature on the topic, this paper proposes a critical perspective, emphasizing that the Blockchain should not be considered as a one-trick tool to be applied to every possible domain but, rather, a new technological advancement with its own niche of use. A similar perspective is also discussed in \cite{Puthal2}. Particular applications of DLTs to store healthcare informations and IDs are presented in \cite{Healthcare} and \cite{PKI}, while \cite{Novo} proposes to employ Blockchain as a means for arbitrating roles and permissions in IoT.\newline 

Despite all these use cases, to the best of the authors' knowledge, the use of DLTs in a control setting have yet remained unexplored: our interest, as anticipated in an earlier section, stems from the possible applications of DLTs in a smart city environment, using the digital tokens as a way to enforce the desired level of compliance in the resource-sharing interactions between humans and machines. A natural question that arises in this context is the value of the token that would be lost when an agent is not compliant. If this value is too high, (economic) activity stops and resources are not fully utilised; if it is too low, then compliance levels will be low and resources will not be effectively (perhaps optimally) utilised.\newline  

It is worth noting that the idea of using control signals in related areas is not new. For example, link pricing concepts are standard in networking \cite{srikant}, and many stochastic signalling strategies can be interpreted as a price \cite{book,arieh,cdc}. More specifically, a large number of papers have been published on the topic of dynamic pricing being used to increase the quality of service provided to customers in various domains. The underlying concept is sometimes referred to as \emph{Transactive control}, which is the design of feedback loops using financial transactions to improve quality of service in domains such as cities or the smart grid area: particular examples on this topic can be found in \cite{Phan}-\cite{Junjie}. Other specific examples of dynamic pricing can be found in \cite{Li} (incentivizing users to schedule electricity-consuming applications more prudently), in \cite{Chekired} (managing  EVs charging and discharging in order to reduce the peak loads), in \cite{Bejestani} (combining the classical hierarchical control in the power grid with market transactions) and \cite{Hao} (where the authors propose a transactive control system of commercial building heating, ventilation, and air-conditioning for demand response).\newline 

In all the aforementioned works, the control mechanism is obtained using economic transactions in the form of \emph{dynamic toll pricing}; what we propose is a form of \emph{dynamic deposit pricing}. The subtle, but substantial difference between the two approaches is that in the first case a user has to pay a certain amount of money in order to access a service, whereas in the latter, a user is forced to deposit a certain amount of tokens that will be returned once certain criteria are met. Thus, in the latter application it is risk that is being priced, rather than a form of demand management. In other words, from the perspective of a single agent, the amount of digital tokens deposited represents a risk for not complying with a set of rules, rather than a price to pay to access a service. It is important to stress that  \emph{dynamic deposit pricing} is not a disguised form of \emph{dynamic toll pricing}: in the first the objective is optimize the use of a resource shared among agents, in the latter the objective is to ensure that the level of compliance is high enough to not jeopardize the system performance (or even lead it to instability). To give a practical example, consider a digital driving license associated with a fixed amount of tokens. Every time the driver crosses a checkpoint (e.g., a junction with traffic light) they deposit a certain amount of tokens and receive them back only if they behave as a ``good citizen" (i.e., they do not break any rule while driving). In this example, the amount of tokens does not represent the amount of money needed to access a service but rather, a way to ensure that a shared resource is accessed fairly by all the agents involved. Section \ref{Sec: Compliance} provides an example to show how low levels of compliance negatively affect the performance of a traffic junction and how this token model can be employed in such a setting. It seems pretty clear then that, while both approaches use a form of dynamic pricing as control signal, the domains of applications are widely different.\newline


The focus of this paper is thus twofold: in the first part we present a concise, accurate, model for a type of DLT dynamics that would be suitable for implementing the kind of dynamic deposit pricing strategies which we have envisaged. The DLT model is designed to address the need for high-frequency, low-latency transactions,
and we analyse its properties, providing details and a theorem about its stability. The second part presents a class of applications based on the use of DLT to regulate compliance levels at interacting junctions in a road network. The situation described is just an instance of a vast class of problems where human users are expected to adhere to a predetermined set of rules in order to ensure an efficient sharing of resources (see Section \ref{Subsec: Cyber-phisical} for a set of examples).\newline 

The contributions of the paper can thus be summarised as follows.\newline

\begin{itemize}
\item A detailed model of a practically important DLT designed for IoT applications: extensive Monte Carlo simulations are provided to validate its behaviour and show the accuracy of the proposed set of equations.\newline
\item An analysis of the stability of the DLT under the hypothesis of a high arrival rate of transactions. Here, as we shall see, stability relates to the ability to double spend tokens (commit fraud).\newline
\item A \emph{dynamic deposit pricing} mechanism to regulate the compliance levels at interconnected activities to ensure efficient access to a shared resource: sufficient conditions are provided for the stability of the system.\newline
\item An application of the pricing token mechanism to road intersections where vehicles are expected to respect the constraints imposed by the traffic light in order to access the junction.\newline
\end{itemize}
\subsection{Organization}
The reminder of the paper is organised as follows. Section \ref{Sec: DLT structures} describes the most well-known DLT structures and provides an analysis of what requirements a cryptocurrency must have to be used in smart-city related compliance problems. Section \ref{Sec: Tangle} describes the mathematical modelling of the Tangle: first a more general and stochastic system is presented and validated, then, under the assumption of a high arrival rate of transactions, a deterministic fluid model is derived and analysed in depth. Section \ref{Sec: Compliance} describes the use of pricing tokens  in smart city compliance problems: examples and simulations are provided to show how low levels of compliance affect negatively the performances of these systems. Finally, main conclusions and future work are outlined in Section \ref{Sec: Conclusions}. Note finally that the literature on DLT architectures is split between academic publications and non-archival sources (such as forums, web pages, and news articles). We adopt the convention that non-archival sources are given in the form of footnotes, and more conventional publications in the reference list.

\section{Discussion of Basic DLT structures}
\label{Sec: DLT structures}
A DLT is a spreadsheet where a record of transactions and other account information is transcribed, accessible and (potentially) owned by every node of a Peer to Peer (P2P) network, with an intrinsic mechanism to enforce consensus among its users.\newline 

{Such systems, despite being conceptually simple, must possess certain properties to be useful in a large-scale compliance problems.\newline
\begin{itemize}
\item[P1] \underline{DLT architectures must be scalable:}. That is,  for IoT applications, the number of transactions per second between devices can be in the order of thousands. Therefore any infrastructure needs to be able to manage such an amount of operations.\newline
\item[P2] \underline{Double-spending: } The DLT infrastructure must be resilient to attacks from malicious users (e.g., the structure must be safe against {\em double spending} attacks). Here, by {\em double spending}, we mean the ability of an agent to spend the same token more than once. \newline
\item[P3] \underline{Energy costs:} With the energy costs of {\em bitcoin} already approaching absurd levels, the energy cost to maintain the network infrastructure consistent and safe from attacks has to be kept at a reasonable level.\newline 
\item [P4] \underline{Fees:} Transactions should be free of transaction costs. This is an extremely important feature from a control perspective. Consider the examples already discussed; if any time a transaction among devices enforces the payment of a fee, this will eventually deplete the coin value, thereby hindering its ability to participate in the network regulation problems\protect\footnote{{Note that, strictly speaking, pure-fee less transactions, in general, are clearly not possible. 
For example, nodes must always expend energy to support the operation of the overall DLT system. 
We distinguish between these shared {\em infrastructure} costs, and the {\em transaction} costs that are paid to a third party to support the validation of transactions. Some DLTs, (such as IOTA - see later) do not insist on such fees. In the case of IOTA, infrastructure costs are shared among community members; namely, when a user transacts, that smae user also validates another users' transaction. This is in contrast to blockchain where transactions are validated by special community members (the miners) who are compensated for their efforts.}}.\newline
\item [P5] \underline{Price volatility:} Trading a fixed number of tokens on open platforms may be subject to significant price fluctuations. Developing economic systems, or using such tokens to implement a control strategy is difficult, due to possible hoarding of tokens. Thus, the ability to create tokens that are fixed against a stable currency, such as the US dollar, is very important.\newline
\end{itemize} }

In this Section we provide a description of two widely used DLT architectures, traditional {\em Blockchain} and the {\em Directed Acyclic Graphs} (DAGs). Note that such systems can be compared from an architectural perspective, from the perspective of the above items, or from the perspective of the consensus mechanism; for example, Blockchain is a competitive consensus system, whereas DAGs typically operate a swarm type of consensus mechanism (that use a type of majority voting algorithm).

\subsection{Blockchain}

Blockchain was first introduced by Satoshi Nakamoto in his seminal white-paper \cite{Nakamoto} as the technology on which the Bitcoin was developed. Since that first paper, and following the success of Bitcoin, a large number of other currencies have been developed trying to emulate or to improve the original design. Almost all these currencies, at their core, share the same functioning architecture
introduced by Nakamoto.\newline 

Blockchain is a peer to peer (P2P) distributed ledger of transactions \cite{Swan}, meaning that the ledger file (i.e., the spreadsheet that holds every transaction record) is not stored at a central server, but rather copies are distributed across a network of private computers (nodes). In order to exchange currency (or information), nodes issue transactions among each other using public/private key cryptography \cite{Puthal}. Every account-holder has a public key and a private (secret) key. The latter is used to sign/authenticate transactions, whereas the public one provides a unique address in the system. 

\begin{ex}For instance, in order for a user Paul to send a certain amount of currency to Anna, he needs to write a transaction, signed with the private key of the address where the coins are stored, attach his own public key to identify himself as the sender, and address it to Anna's public key so as to identify her as the receiver. The transaction states that Paul's account balance will decrease by some amount and that Anna's account balance will increase by the same or lesser amount (any difference will be taken as a transaction fee by the successful miner, see below). This transaction is broadcast to the network and, after validation, is applied to every copy of the ledger.\end{ex} 

\begin{figure}
\includegraphics[width=1\columnwidth]{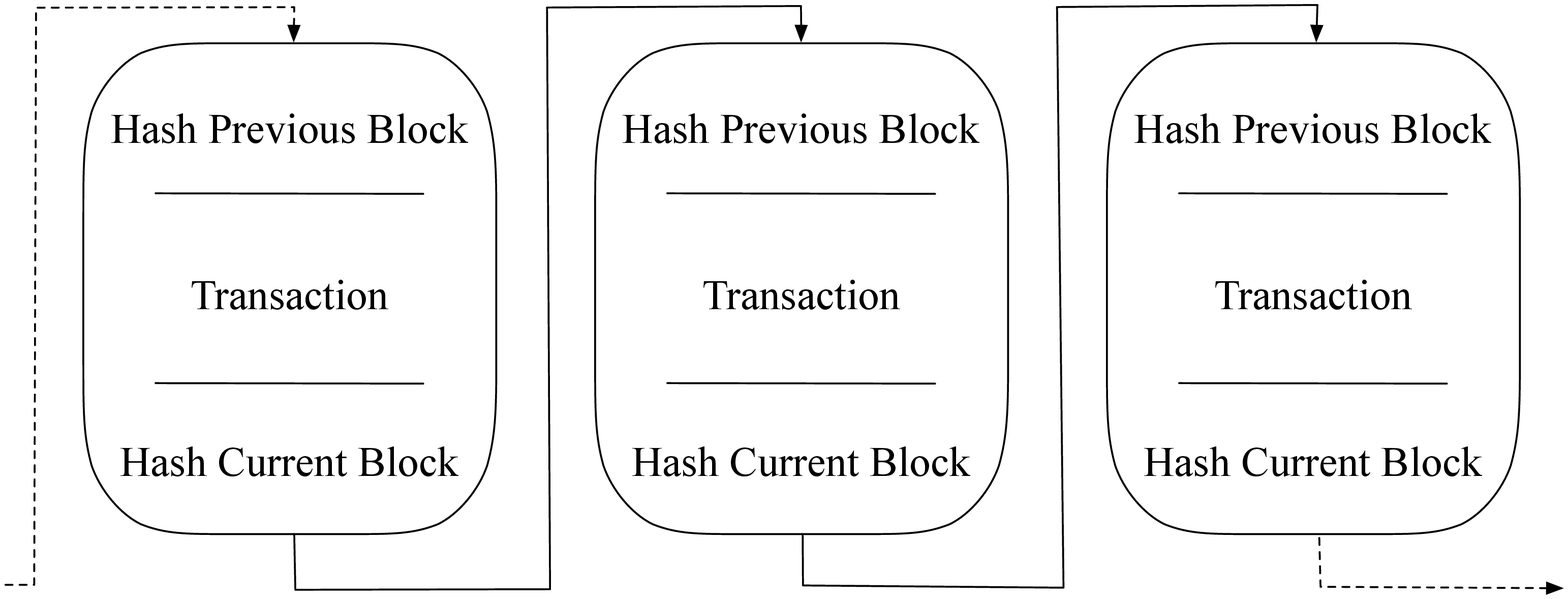}
\caption{Visual representation of three blocks of a Blockchain. Each block references the previous one and therefore any change in any part of the chain would result in an inconsistency.}
\label{Fig: Blockchain}
\end{figure}

To be validated, transactions are sent to the P2P network.  If they are consistent with all previous transactions,  some specific nodes called miners collect them together into data structures called blocks (each miner can decide which transactions to add to a specific block).  Once a new block is proposed by a miner, it is sent to the network; if it is valid, every other node will add it to the end of the Blockchain. Each block contains a reference to the previous block, and transactions in the same block are considered to have happened at the same time; refer to Figure \ref{Fig: Blockchain} for a visual representation of this process. In order to  incentivise nodes to become miners, for each transaction approved, a miner is rewarded by some amount of currency, in the form of a transaction fee or bitcoins; in other words, each user needs to pay a miner a certain amount of currency for the service of adding blocks to the Blockchain.\newline 

Assuming consistency of transactions, miners must complete a certain amount of work to validate a new block in the Blockchain. This 
mechanism is known as {\em Proof of Work} (PoW). PoW involves solving a computationally-hard puzzle; specifically,  the node that performs it needs to operate brute force computations to find a particular \emph{hash} (the image of a hashing function \cite{Hash}) in a high dimensional space that satisfies certain conditions (due to the nature of the problem, it is not feasible to perform anything different than a brute force approach). The miner which adds the next block to the Blockchain is the first one who is able to compute a valid hash and hence solve the puzzle. The amount of computations needed to solve it is very high and as long as the total computational power of the honest nodes is greater than the computational power of attackers that try to perform double spending transactions, honest nodes will outpace dishonest ones and only legitimate transactions will be part of the Blockchain. {In other words, any malicious attempt to tamper with the Blockchain ledger (e.g. trying to alter informations on past transactions) can succeed only in the presence of a sufficient amount of computational power. Therefore the security of the Blockchain resides on the current amount of hashing power that is being used to validate transactions in the network}. For more information regarding the security of Blockchain systems, the interested reader can refer to \cite{Blockchain Security 1}-\cite{Blockchain Security 4}. In the event two or more miners solve the PoW at the same time, to avoid conflicts, the Blockchain system requires each node to build immediately on the longest chain available. In other words, if two valid blocks are issued at the same time, they will both be accepted in the Blockchain: at this stage there will be two or more possible chains on which miners can try to add further blocks. As soon as one block is added to one of the chains, this one (the longest chain) will be accepted by the network and all the other ones will be discarded. In this way, the Blockchain enforces consensus and avoids possible forks that might endanger consistency. Thus, Blockchain may be thought of as a system that is built on a {\em competitive} consensus method. \newline 

Despite its simplistic brilliance, PoW suffers from a major drawback that greatly compromises its use in the long term: in order to keep the network secure, the amount of energy consumption to perform the computations for the PoW is tremendously large. In order to solve this issue a system called {\em Proof of Stake} (PoS), based not on the amount of computational power employed but on the basis of the amount of currency owned by each miner, has been developed. In PoS, a validator, who is the equivalent of a miner in the PoW is elected to add a further block to the Blockchain, on the basis of the amount of currency possessed, and depending on the age of the coins (usually called the maturity date). A user with more substantial amount of coins held for a long time will be more likely to validate a block. As in the PoW, the validator is rewarded with a transaction fee for their effort. This mechanism requires little or no computational power to be executed and therefore the energy costs involved in the PoS are negligible compared to the PoW. The interested reader can refer to \cite{Ouroboros} for a PoS-based Blockchain protocol.\newline 	

Finally, it is worth noting that the Blockchain presents scalability issues in the form of transaction per second: at the current time, the two most famous Blockchain based cryptocurrencies, Ethereum and Bitcoin, are able to process up to 20 and 7 transactions per second. This, in addition to the fact that transaction fees appear to be necessary as an incentive for transaction approvals, poses the question of whether or not the Blockchain is a good candidate DLT in a compliance enforcement scenario.

\subsection{Directed Acyclic Graphs}
\label{Sec: DAG}
\begin{figure}
\includegraphics[width=1\columnwidth]{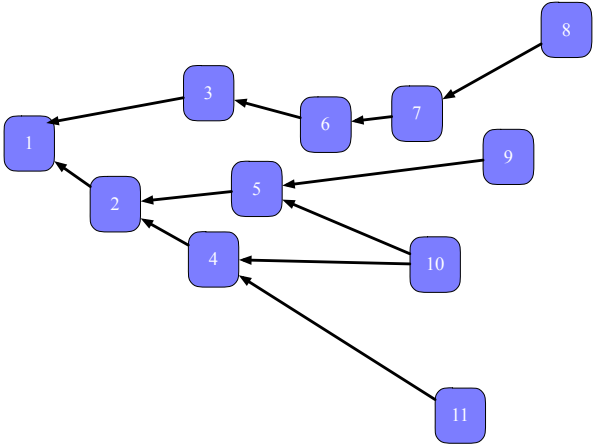}
\caption{Example of a DAG with 11 vertices and 10 edges. All the vertices are directed and it is impossible to find a path that connects any vertex with itself.}
\label{Fig: DAG}
\end{figure}

A different solution for achieving consensus in a DLT uses a Directed Acyclic Graph (DAG).  The DAG is the backbone of several cryptocurrencies (e.g., IOTA\footnote{https://www.iota.org}, Byteball\footnote{https://byteball.org}, Nano\footnote{https://nano.org}). More specifically, a DAG is a finite directed graph with no directed cycles. In other words, it is a graph that consists of a finite number of vertices and edges, with each edge directed from one vertex to another, such that there is no path that connects a vertex $\nu$ with itself. An example of a DAG is depicted in Figure \ref{Fig: DAG}. {  Notice that technically the graph depicted in Figure  \ref{Fig: DAG} contains one cycle (i.e. vertices 2, 4, 5, 10), but the edges connecting one vertex to another are directed in only one direction, therefore making it impossible to find a path between any vertex to itself.}   \newline 

\begin{figure}
\includegraphics[width=1\columnwidth]{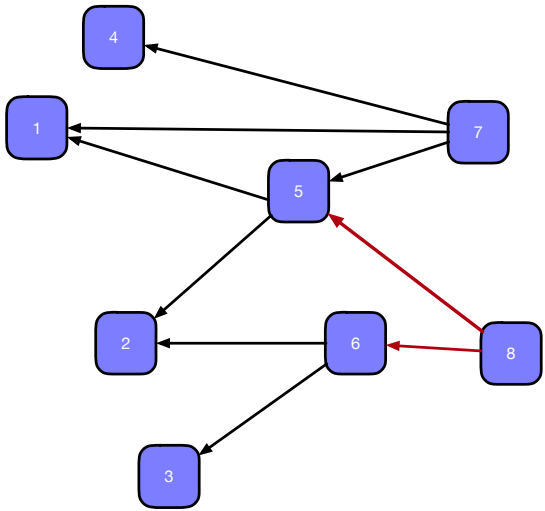}
\caption{Transaction 8 directly approves 5 and 6. It indirectly approves 1, 2 and 3. It does not approve 4 and 7.}
\label{Fig: DirUndir Approval}
\end{figure}

A particular instance of a DAG is the IOTA Tangle \cite{Popov}. Here the consensus method is cooperative in nature, rather than competitive as is the case in Blockchain. The objective of the Tangle, according to the original paper, is to design a cryptocurrency for the IoT industry with its main features being the absence of fees and low energy consumption. The Tangle is basically a DAG where each vertex represents a transaction, called a \emph{site} (in the rest of the paper we might use either site or transaction to refer to a vertex of the Tangle), whereas the  graph represents the ledger. Whenever a new transaction is issued, this must approve two previous transactions. Each approval represents an edge of the graph. All yet unapproved sites are called \emph{tips} and the set  of all unapproved transactions is called the \emph{tips set}. A directed edge between site $i$ and site $j$, means that $i$ directly approves $j$, whereas, if there is a path (but not a single edge) that connects $i$ to $j$, we say that  $j$ is indirectly approved by $i$ (e.g.,  see Figure \ref{Fig: DirUndir Approval}). The first transaction in the Tangle is called the \emph{genesis} block (i.e., the transaction where all the tokens were sent from the original account to all the other accounts) and all transactions indirectly approve it. Furthermore, in order to prevent malicious users from spamming the network, whenever a new transaction undergoes the approval step, it has to perform a PoW (a much lighter version than the one performed by the Blockchain). In what follows, we assume that there is a simple way to verify if, during the approval phase, the selected transactions are consistent among each other and with all the sites directly or indirectly approved by them. In case they were not, the selection process needs to be run again, until two consistent transactions are found.\newline

\begin{figure}
\includegraphics[width=1\columnwidth]{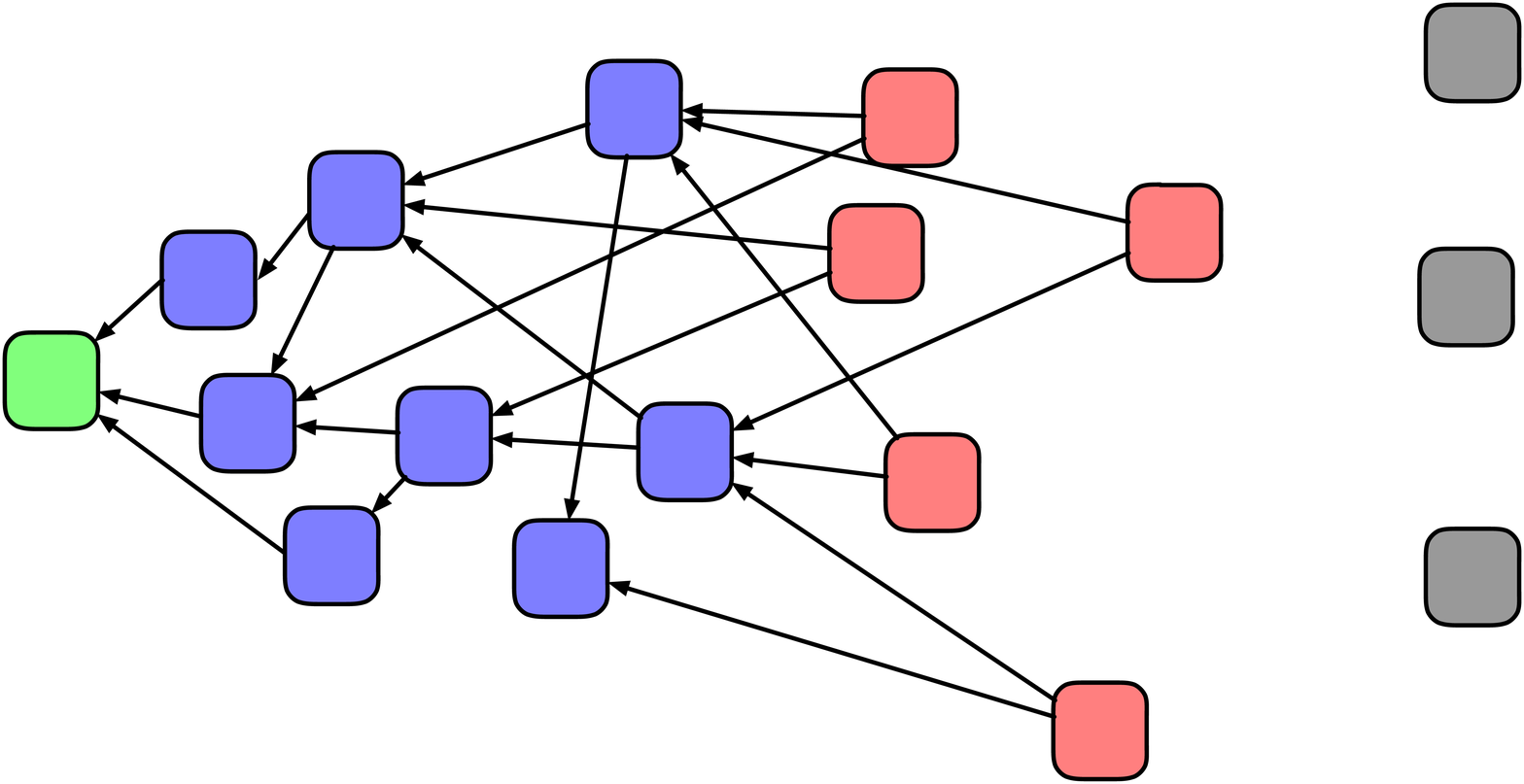}
\includegraphics[width=1\columnwidth]{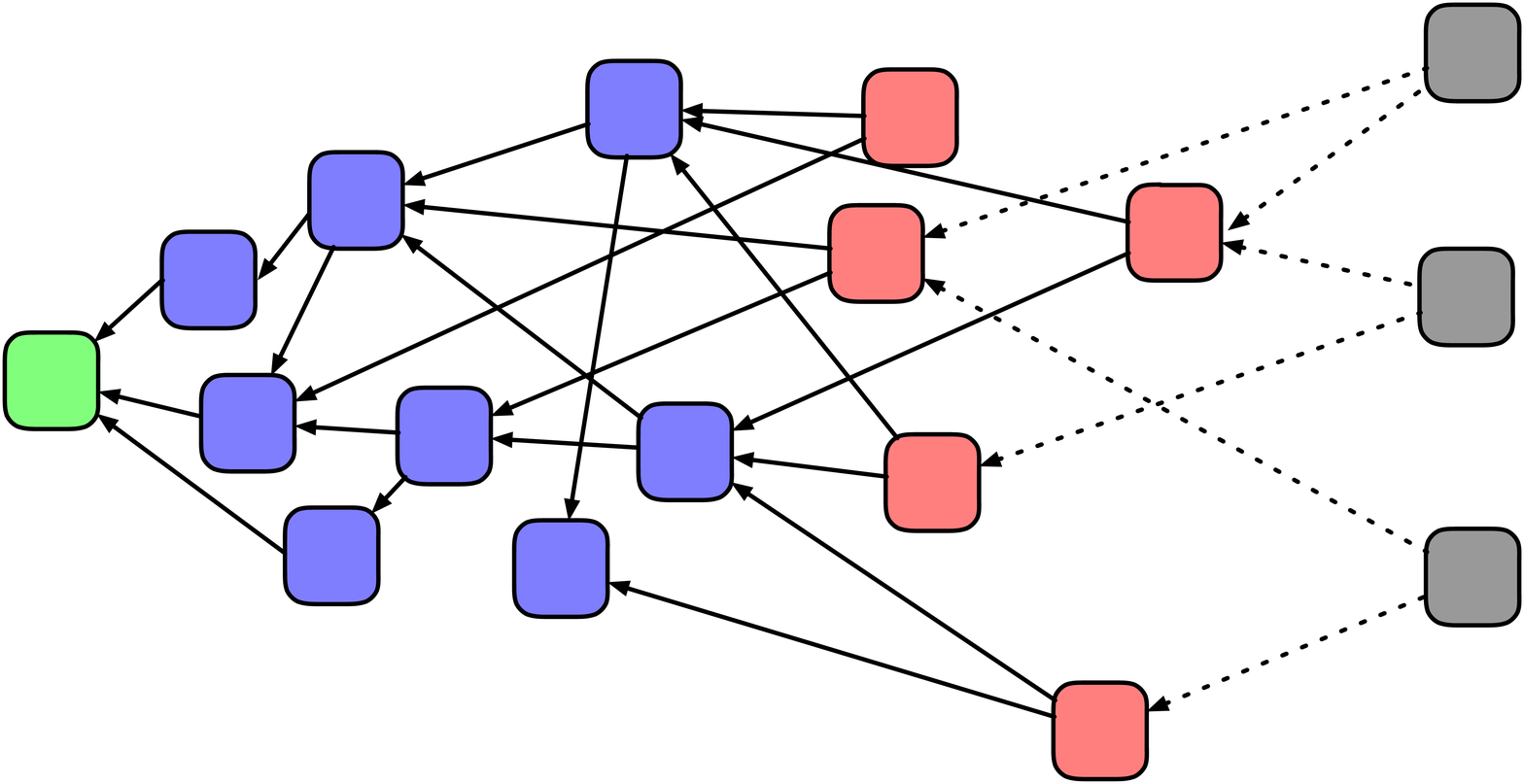}
\includegraphics[width=1\columnwidth]{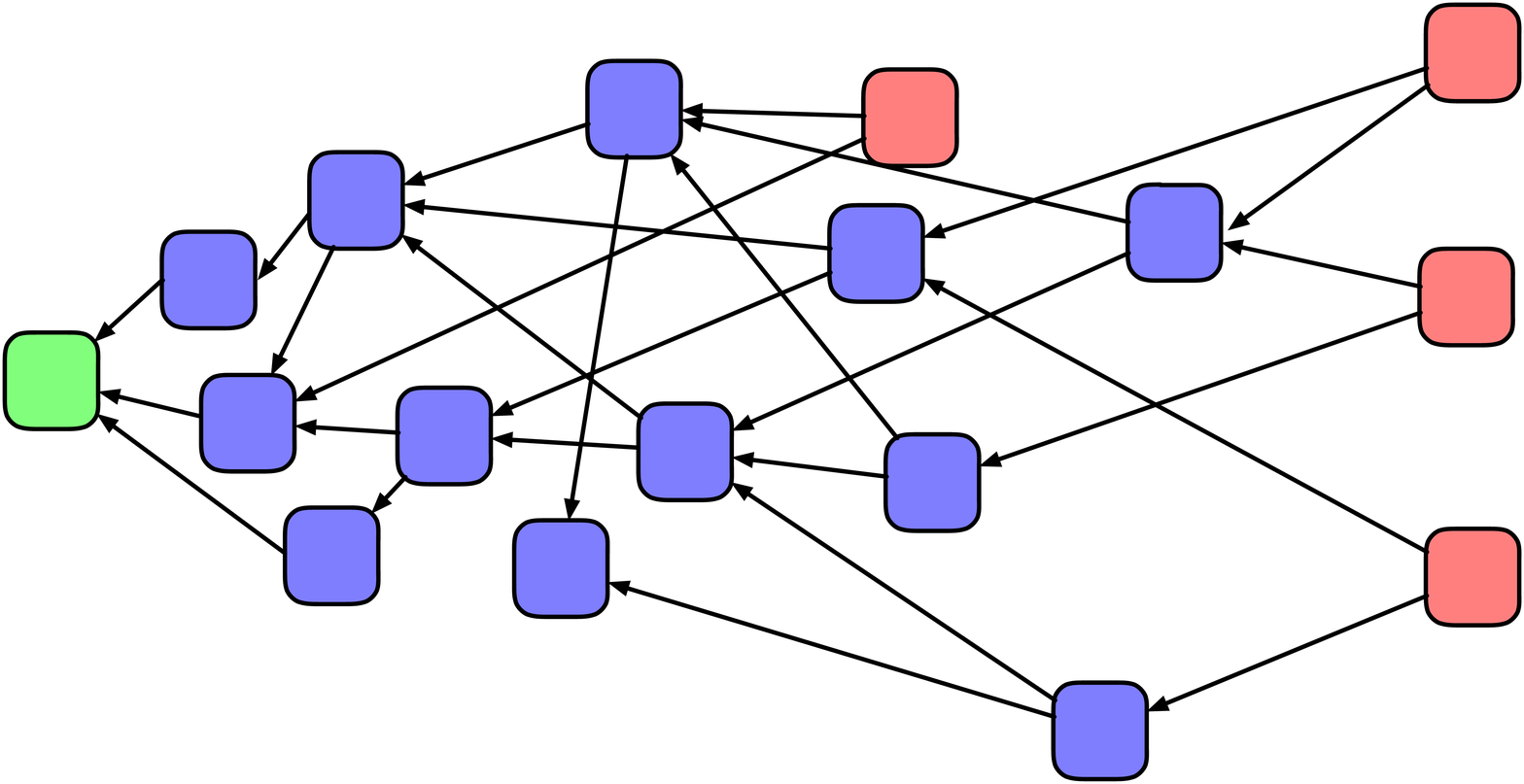}
\caption{Sequence to issue a new transaction. The green site represents the genesis block, the blue sites represent the approved transactions and the red ones represent the tips. The back edges represent approvals, whereas the dashed ones represent transactions that are performing the PoW in order to approve two tips.}
\label{Fig: Tangle}
\end{figure}

To have a better understanding of the previous process refer to Figure \ref{Fig: Tangle}: a certain instance of the Tangle, with three new incoming transactions is presented (upper panel). The green block (the leftmost) is the genesis site, blue blocks are transactions that have already been approved, red blocks represent the current tips of the Tangle and grey blocks are new incoming transactions. Immediately, when  issued,  a new transaction tries to attach itself to two of the network current tips (middle panel). If any of the selected tips was inconsistent with the previous transactions, or with each other the selection would be rejected and two other tips would be selected. Notice that at this stage, these transactions are not yet part of the Tangle as they are carrying out the required PoW and the tips remain unconfirmed (dashed lines) until this process is over. Once the PoW is finished, the selected tips become confirmed sites and the grey blocks are added to the tips set (lower panel). 

\begin{figure}
\includegraphics[width=1\columnwidth]{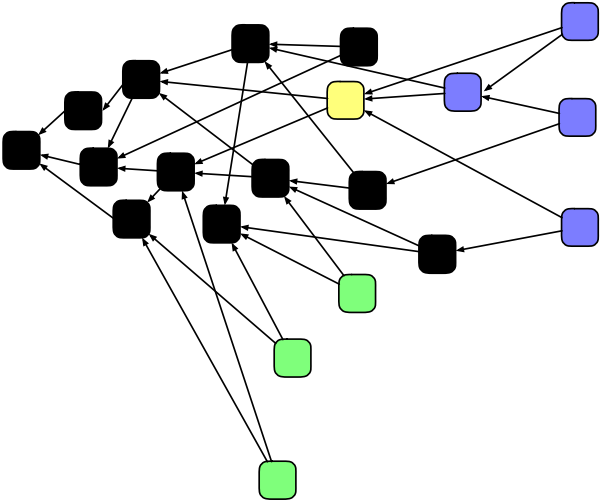}

\caption{The blue and the green transactions are incompatible with each other.}
\label{Fig: DoubleSpending}
\end{figure}

Let us consider another example to explain with more detail the process of approval. Figure \ref{Fig: DoubleSpending} shows a further instance of the Tangle. A malicious user sent a certain amount of money to a merchant. The corresponding transaction is the yellow block in the figure. The same user, afterwards, makes other transactions trying to double spend the same money that were sent to the merchant. These correspond to the green blocks. It is worth stressing, at this point, that there is no mechanism to force a user to select certain sites for confirmation. Any transaction can be selected as long as it is consistent with the sites that are approved (directly or indirectly) by it. Nevertheless it is reasonable to assume that the vast majority of nodes would have little interest in confirming specific transactions and would follow the tips selection algorithm proposed by the protocol. In this scenario, all the transactions that approve the original yellow site (the blue blocks) are incompatible with the green ones, therefore any new transactions can either approve the green/black sites or the blue/black ones. The green/blue combination would be considered invalid (as there is an inconsistency in the ledger) and a new selection would be made. The objective of an hypothetical attacker would be then to wait for the merchant to accept their payment, receive their goods, then create one or more double spending transactions that get approved by other honest sites. If this is the case, what prevents nodes from spending their money twice? Roughly speaking, depending on the tip selection algorithm employed, due to the presence of the PoW, any malicious user would need to possess the majority of the hashing power in the Tangle  to perform such an attack and make it successful. A thorough discussion on the possible attack scenarios of the Tangle is beyond the scope of this paper and we refer the interested reader to \cite{Popov} and \cite{Bramas}.


\subsection{Blockchain vs. DAG: A social compliance perspective}
While Blockchain is without doubt the genesis of the current interest in DLT, its use in IoT scenario is limited: {  apart from the large energy costs, and the volatility of coins such as bitcoin, the long transaction times,
transaction fees, and the inherent preference for miners to process large transactions rather than small ones, makes 
its use of limited value and represent a bottle neck for its adoption in a scenario where thousands of devices communicate with each other many times per second.  In contrast, the DAG structure, when used to create non-traded tokens (to remove price volatility - see P5 above),  may be much more suitable 
for such use cases as it is claimed to support high-frequency (low latency) micro trading.}
Thus, in the remainder of this paper we focus solely on a mathematical description of the DAG as envisaged in the Tangle, and explain some
 use cases that relate to compliance in smart cities.

\section{The mathematics of the Directed Acyclic Graph}
\label{Sec: Tangle}
A {\em directed acyclic graph} (DAG) $G=(V,E)$ consists of a set of vertices or sites $V$ and a set of directed edges $E$, with the additional condition
that the graph has no cycles (a cycle is a collection of directed edges $\{ e_i =  \la u_i, v_i \ra \in E \,:\, i=1,\dots,k\}$
such that $v_i = u_{i+1}$ for $i=1,\dots,k-1$ and $v_k = u_1$). 
When an edge $e = \la u,v \ra \in E$ is directed from $u$ to $v$ (where $u,v \in V$) we will say that $v$ is a parent of $u$,
and that $u$ is a child of $v$.
Viewed at a high level, a Blockchain is a simple example of a DAG,
where the site set $V$ consists of the blocks, and the edge set $E$ consists of the directed edges from each block to its
unique predecessor in the chain (except for the genesis block which does not have a predecessor). 
In our applications the DAG will be connected, and as in the Blockchain there will be a unique site (the genesis)
which has no parent sites. In a finite DAG the acyclic condition implies that there must be at least one site with no children
(this can be seen by following a path which always moves in the direction from parent to child).
The sites with no children are referred to as the {\em tips} of the graph.

\subsection{The IOTA Tangle}
The Tangle is an increasing family of finite DAG's $\{G(t) : t \ge 0\}$ where each site of $G(t)$
contains the record of a transaction which arrived at or before time $t$. The DAG
grows through the addition of new sites which represent newly arrived transactions. The special feature of the Tangle is that each new site
is attached to two pre-existing sites on the graph by directed edges, meaning that the new site becomes the child of both of these pre-existing sites.
Furthermore both of these pre-existing sites must be tips of the graph at the time when they are selected by the new transaction.
There is a delay between the time when the tips are selected and the time when the new site is attached to the Tangle.
This delay allows time for a proof of work and for the validation of the transactions in the two parent sites (this validation ensures that the transactions on
the selected tips are consistent with each other and with their parent sites). Hence the
directed edges from the new site represent approvals of the transactions which reside at the existing parent sites
(note that these approvals do not imply any relation between the transaction in the new site and the transactions in the two parent sites).
The weight of a site is (one plus) the number of sites which have approved it either directly or indirectly, which is also (one plus) the number of its descendants
in the graph.
Thus the weight of a site represents the amount of work which would be needed to repair the Tangle if the transaction represented by that site were altered,
and so the weight measures the security of a transaction in the Tangle (in the Blockchain the analogous measure of security
for a transaction is the accumulated difficulty of blocks which have been inserted subsequent to its own block).

\medskip
We will describe a mathematical model for the growth of the Tangle $G(t)$ in a situation where one central user maintains the
record of the Tangle, and other users generate transactions for inclusion in the Tangle. In a real network many users would maintain
local copies of the Tangle, and each user would independently update its own copy. 
However this complicates the analysis since it can lead to synchronization issues for updates.
Therefore we assume the simpler scenario where only one user maintains the record, and other users view this record when they
create a new transaction.
The model will depend on
(i) an arrival process $N(t)$ which describes the creation of new transactions, 
(ii) a tip selection algorithm, which will describe how each newly created transaction selects two tips for approval, and
(iii) a delay time $h$ which accounts for the time between creation of a transaction and its addition to the Tangle (for reasons of simplicity $h$ will be assumed
to be the same for all transactions).

\subsection{Random tip selection model}
\label{Sec: Simple model}
Our analysis will focus on the simplest tip selection algorithm, in which each new transaction randomly and independently selects two
tips for approval. 
We call this the {\em random tip selection algorithm}, and we will analyze the growth of the Tangle under this assumption.
Keeping track of the growth of the full DAG would be quite complicated, but fortunately
the random tip selection algorithm allows us to construct a simpler model which just keeps track of the number of tips.
In this setting the growth of the Tangle is determined by the sequence of times when new
transactions are created, and by the selections of tips for approval. 
The  random tip selection algorithm was introduced and discussed in the paper \cite{Popov}, and later we will compare 
the predictions of our model with some of the results derived there.

\medskip
We assume that when a transaction is created
two tips are immediately selected and validation is attempted 
(note that the same tip may be selected twice by the new transaction, since in the random tip selection algorithm 
the two choices are made independently). If validation fails
the choices are discarded and another
two tips are selected for validation. This continues until the process is successful, and we assume that this whole
validation effort is essentially instantaneous. However after the validation there is a waiting period $h$  during which
the proof of work is carried out and the transaction is communicated to the central user where the Tangle is stored. 
During this time the approvals of the selected tips are pending, so the tips may still be
available for selection by other new transactions. After the waiting time $h$ the two new directed edges are
added to the graph, directed from the new site to its two parent sites. After this point the two parent sites
are no longer tips, and so are no longer available for selection by other new transactions
(note that the parent sites may have been previously selected by earlier arrivals, in which case they ceased being tips at an earlier
time).

\medskip
\noindent The reduced model involves these variables:
\begin{itemize}
\item[1)] $L(t)$ is the number of tips at time $t$
\item[2)] $W(t)$ is the number of `pending' tips at time $t$ which are being considered for approval by some new transaction
\item[3)] $X(t) = L(t) - W(t)$ is the number of `free' tips at time $t$
\item[4)] $T_a$ is the time when transaction $a$ is created
\item[5)] $N(t)$ is the number of transactions created up to time $t$
\item[6)] $U(T_a) \in \{0,1,2\}$ is the number of free tips selected for approval by transaction $a$ at time $T_a$
\end{itemize}
We have the relation
\be\label{eqn:N}
N(t) =  \sum_{a : T_a \le t} \, 1
\ee
and similar expressions for the other variables:
\be
W(t) &=& \sum_{a: t-h < T_a \le t} U(T_a) \label{eqn:W} \\
X(t) &=& N(t -h) - \sum_{a : T_a \le t} U(T_a) \label{eqn:X} \\
L(t) &=& N(t-h) - \sum_{a : T_a \le t - h} U(T_a) \label{eqn:L}
\ee

\begin{figure}
\includegraphics[width=1\columnwidth]{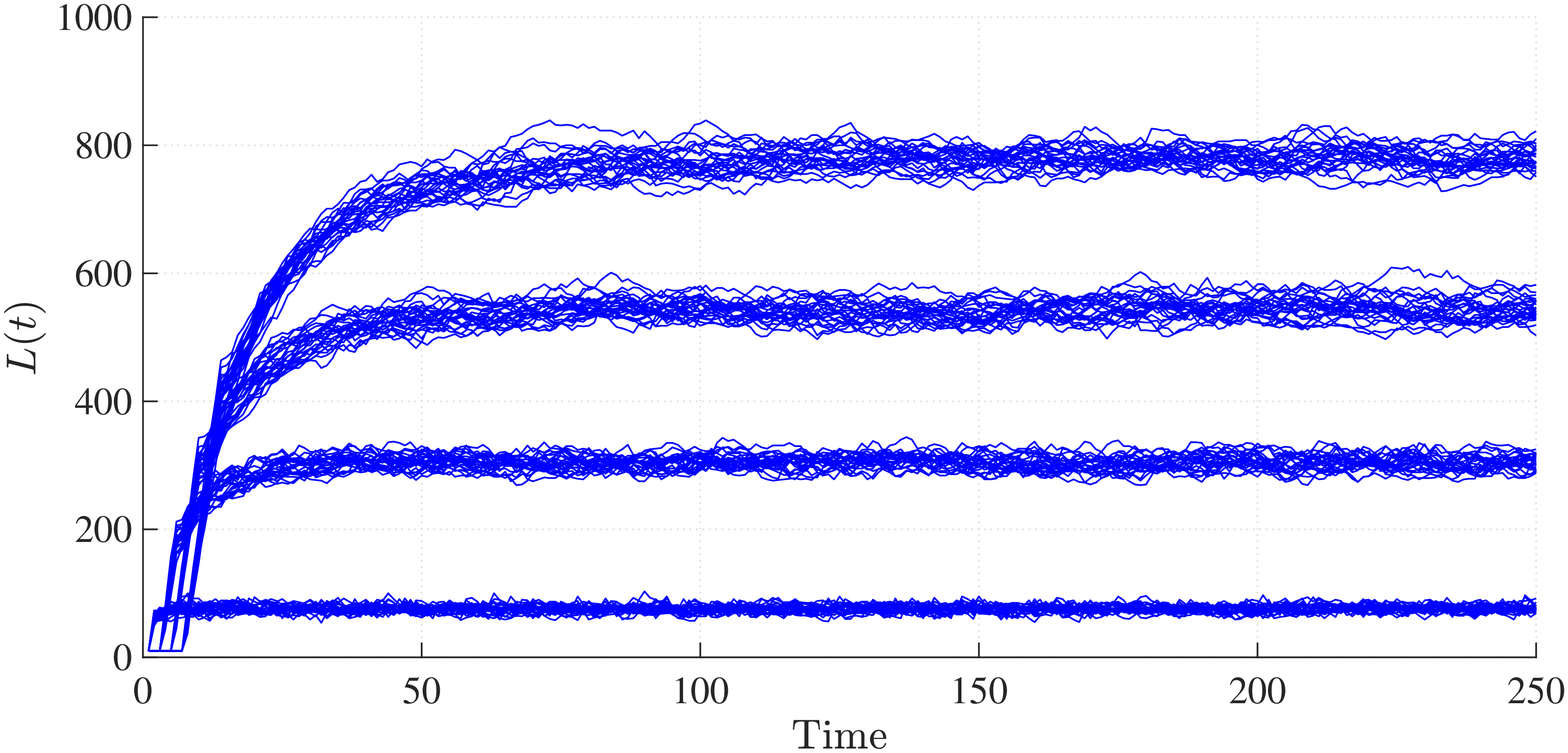}
\includegraphics[width=1\columnwidth]{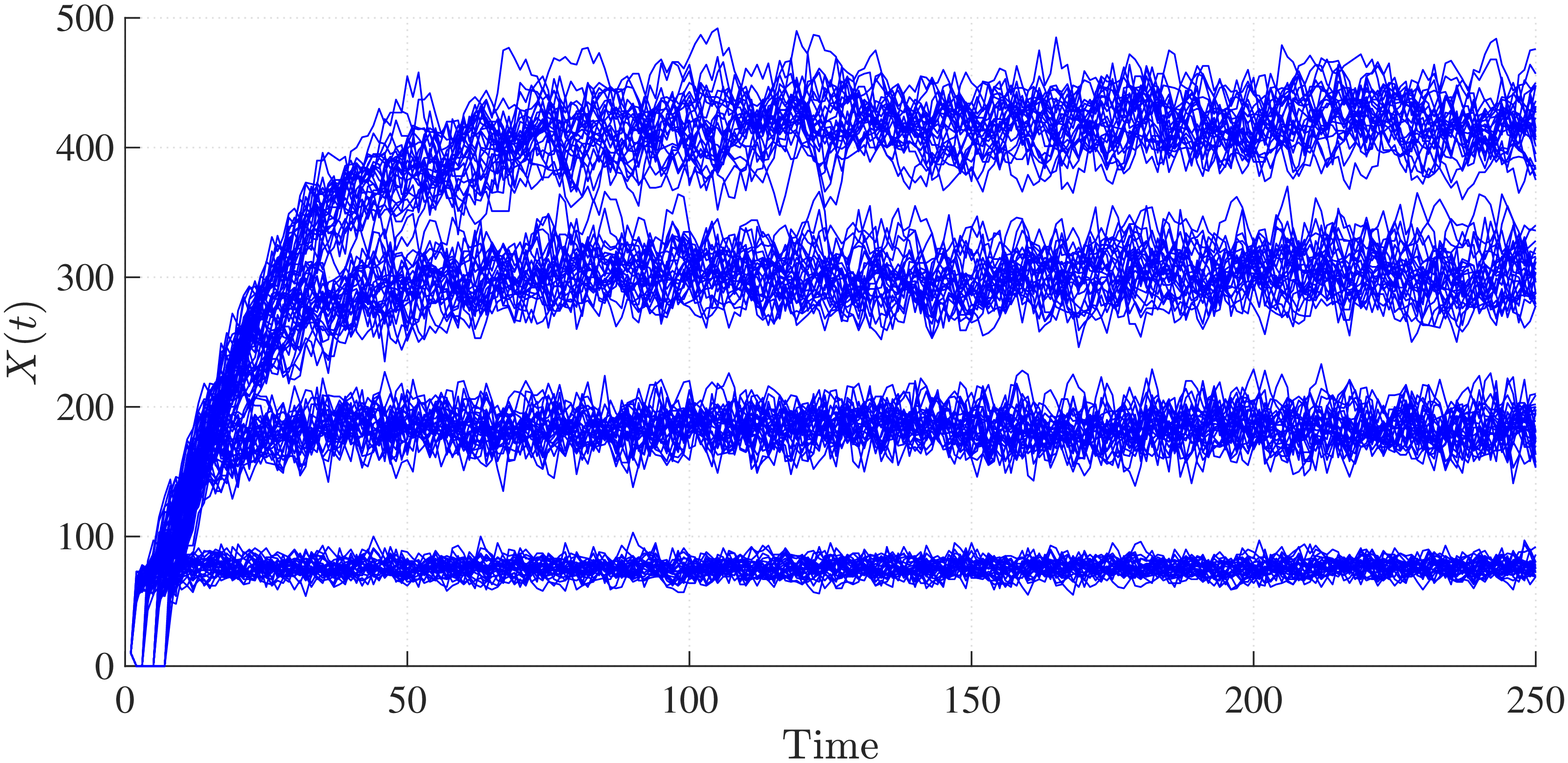}

\caption{Four different simulations of the Tangle with increasing delays, $h \in \{1, 3, 5, 7\}$ (the larger the delay, the larger the steady value). 
Each simulation shows 100 realizations of the reduced model with transaction times, $T_a$, generated according to a Poisson distribution with $\lambda = 60$. 
The upper panel shows the leaves, $L(t)$, whereas the lower panel shows the free leaves, $X(t)$}
\label{Fig: Tangle Discrete}
\end{figure}

\medskip
Assuming the random tip selection algorithm,
$U(T_a)$ is a random variable whose distribution depends only on the values
of $X$, $W$ and $L$ just prior to time $T_a$. We assume that $X,W,L$ have left limits, and
define
$X(t-0) = \lim_{\epsilon \downarrow 0} X(t - \epsilon)$ and similarly for
$W(t-0)$, $L(t-0)$.
Then the distribution is
\be
\P(U(T_a) = 0) &=& \left(\frac{W(T_a - 0)}{L(T_a - 0)}\right)^2 \\
\P(U(T_a) = 1) &=&  \frac{\left(2 W(T_a - 0) + 1 \right) X(T_a - 0)}{L(T_a - 0)^2} \\
\P(U(T_a) = 2) &=& \frac{X(T_a - 0)^2 - X(T_a - 0)}{L(T_a - 0)^2}
\ee
and the expected value is
\be
\E[U(T_a)] = 2 \, \frac{X(T_a - 0)}{L(T_a - 0)} -  \frac{X(T_a - 0)}{L(T_a - 0)^2}
\ee
The system of equations  (\ref{eqn:N}-\ref{eqn:L}) is well-suited for simulation: if the transaction times
$\{T_a\}$ are generated according to some rule  (for example a Poisson process) then 
(\ref{eqn:N}-\ref{eqn:L}) provide the update rules for the variables $(N,X,L,W)$ (updates occur 
at both the creation times $\{T_a\}$ and the times $\{T_a+h\}$ when sites are added to the DAG).
Some examples of the resulting simulations are shown in Fig. \ref{Fig: Tangle Discrete}.
Note that in the paper \cite{Popov} the steady state (average) value of $L$ was 
predicted to be $L_0 = 2 \lambda h$, which is consistent with the simulations shown in Figure \ref{Fig: Tangle Discrete} 
for large arrival rates and large values of $h$.

Although we do not pursue these questions here, it would be interesting to consider 
general properties of the stochastic process defined by (\ref{eqn:N}-\ref{eqn:L}) such as the existence of a stationary distribution and ergodicity,
under reasonable assumptions for $N(t)$ (for example that $N$ is a Poisson process).

\subsection{Conflicts on the Tangle}
\label{Sec: Conflicts}
The Tangle is designed to be a secure, reliable and robust ledger for storing transaction records.
However as explained before it is possible for conflicting transactions to be added to the Tangle at more or less the same time,
and so the only way for the Tangle to achieve its goal of reliable record-keeping is to hope
that conflicting records will be eliminated as the Tangle grows. 
In order to explore this avenue for conflict resolution we consider the situation where
several conflicting sub-DAGs exist on the Tangle, and analyze how the Tangle grows in the presence of
such conflicting records. Recall that a newly created transaction will approve two parent sites that belong
to the same sub-DAG, but will never approve two parent sites 
which belong to different sub-DAGs. Thus each sub-DAG will continue to grow, but the sub-DAGs will never be joined.
If this situation were to continue indefinitely then the Tangle would remain inconsistent, and thus would not serve its primary
purpose as a reliable record of authenticated transactions. This motivates investigating whether the attachment algorithm
can lead to some kind of stochastic forcing that will eliminate all but one of the sub-DAG's.

\medskip
We will examine this question using the random tip selection algorithm previously discussed, and assume
that $d$ conflicting sub-DAGs exist on the Tangle.
Thus every site has a label from the set $1,\dots,d$, indicating the sub-DAG to which it belongs. We will call this
the {\em type} of the site.
As before we assume that when a transaction is created the validation process is started,
and that it ends when two tips of the same type have been chosen.
At this point the site for the new transaction is labeled the same type as the selected tips.

\medskip
The new model involves these variables:
\begin{itemize}
\item[1)] $L_i(t)$ is the number of tips of type $i$ at time $t$, so $L(t) = \sum_{i=1}^d L_i(t)$
\item[2)] $W_i(t)$ is the number of `pending' tips of type $i$ at time $t$
\item[3)] $X_i(t) = L_i(t) - W_i(t)$ is the number of `free' tips of type $i$ at time $t$
\item[4)] $T_a$ is the time when transaction $a$ is created
\item[5)] $\tau_a \in \{1,\dots,d\}$ is the type of the transaction $a$
\item[6)] $N_i(t)$ is the number of transactions of type $i$ created up to time $t$
\item[7)] $U(T_a) \in \{0,1,2\}$ is the number of free tips selected for approval by transaction $a$ at time $T_a$
\end{itemize}

\noindent The probability for a transaction at time $t$ to select two tips of type $i$ at the first attempt is
\be
\P(\text{type  $i$ at first attempt}) = \frac{L_i(t - 0)^2}{L(t - 0)^2}
\ee
If tips of different type are selected, the choice is rejected and a new selection is made. 
This continues until two tips of the same type are chosen.
Thus the eventual selection is conditioned on the event that the two selected tips have the same type, which occurs with probability
$\sum_{i=1}^d \P(\text{type  $i$})$. Hence the probability that a transaction is type $i$ is
\be\label{prob0}
\P(\tau_a=i ) =  \frac{L_i(T_a - 0)^2}{\sum_{j=1}^d L_j(T_a - 0)^2}
\ee
Conditioned on the site being type $i$, the distribution of the random variable $U(T_a)$ is
\bee
&\hskip-0.3in&\P\left(U(T_a) = u \,|\, \tau_a = i\right)  \\
&\hskip-0.3in& = \begin{cases} W_i(T_a - 0)^2 \, L_i(T_a - 0)^{-2}  & u =0 \\
\left(2 \, W_i(T_a - 0) + 1 \right) \, X_i(T_a - 0) \, L_i(T_a - 0)^{-2} & u =1 \\
\left(X_i(T_a - 0)^2 - X_i(T_a - 0) \right)  \, L_i(T_a - 0)^{-2} & u=2 \end{cases}
\eee
and therefore
\be\label{av-Ui}
\E[U(T_a) = u \,|\, \tau_a = i] = \frac{2 X_i(T_a - 0)}{L_i(T_a - 0)} - \frac{X_i(T_a - 0)}{L_i(T_a - 0)^2}
\ee

\medskip
The relations for the variables $(N_i,W_i,X_i,L_i)$ are similar to before: 
\be
N_i(t) &=&  \sum_{a : T_a \le t, \, \tau_a=i} \, 1  \label{eqn:Ni} \\
W_i(t) &=& \sum_{a: t-h < T_a \le t, \, \tau_a=i} U(T_a) \label{eqn:Wi} \\
X_i(t) &=& N_i(t -h) - \sum_{a : T_a \le t, \, \tau_a=i} U(T_a) \label{eqn:Xi} \\
L_i(t) &=& N_i(t-h) - \sum_{a : T_a \le t - h, \, \tau_a=i} U(T_a) \label{eqn:Li}
\ee
The system (\ref{eqn:Ni}-\ref{eqn:Li}) can be used to generate simulations of the growth of the Tangle
in the presence of conflicts. The only change from before is that
at each time $T_a$ when a new transaction is created, the type of the transaction is selected using the
distribution (\ref{prob0}). The variables can then be updated at times $\{T_a, T_a + h\}$ according to the
formulas (\ref{eqn:Ni}-\ref{eqn:Li}).

\subsection{Validation of the Tangle model}
\begin{figure}
\includegraphics[width=1\columnwidth]{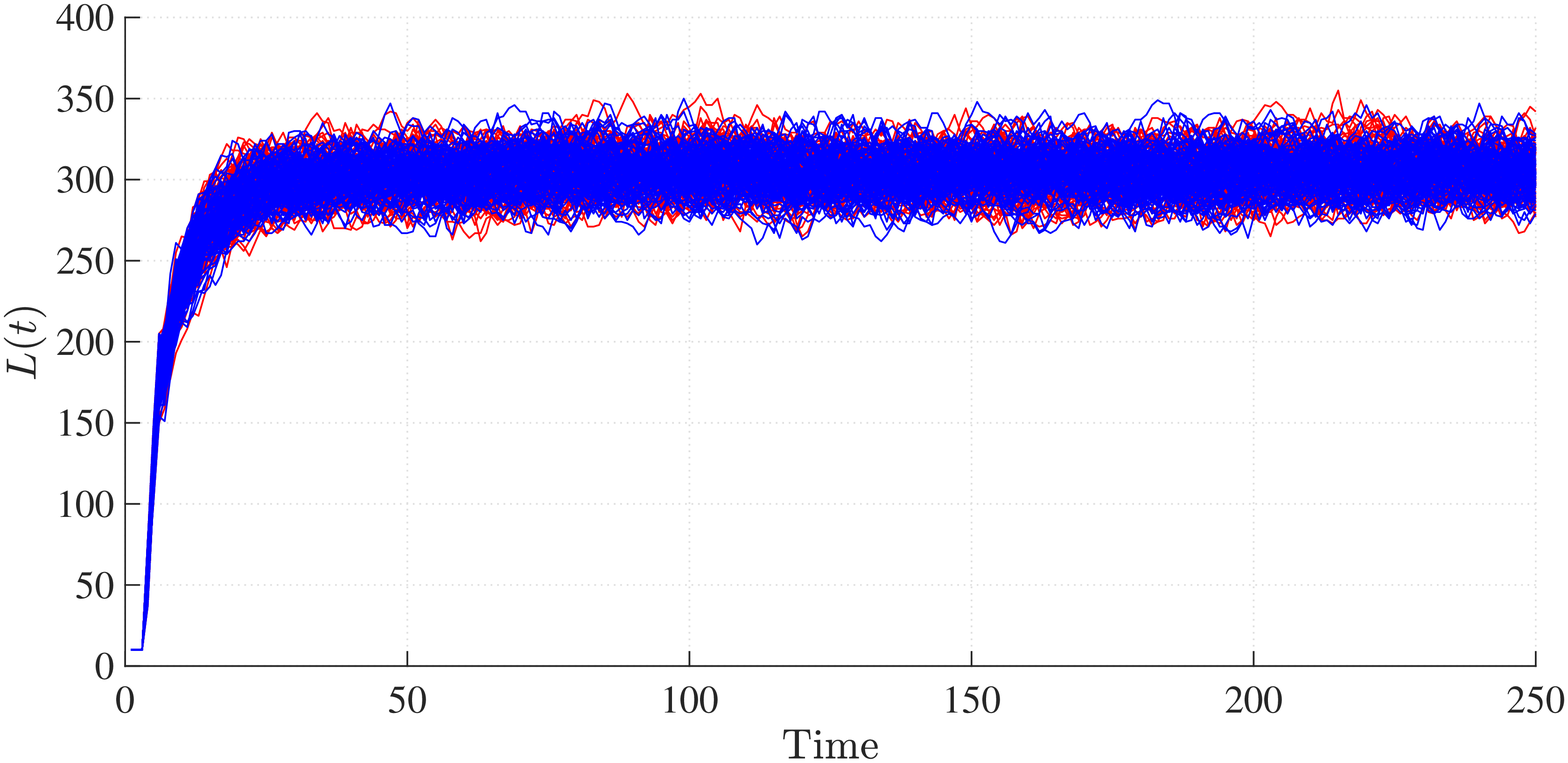}
\includegraphics[width=1\columnwidth]{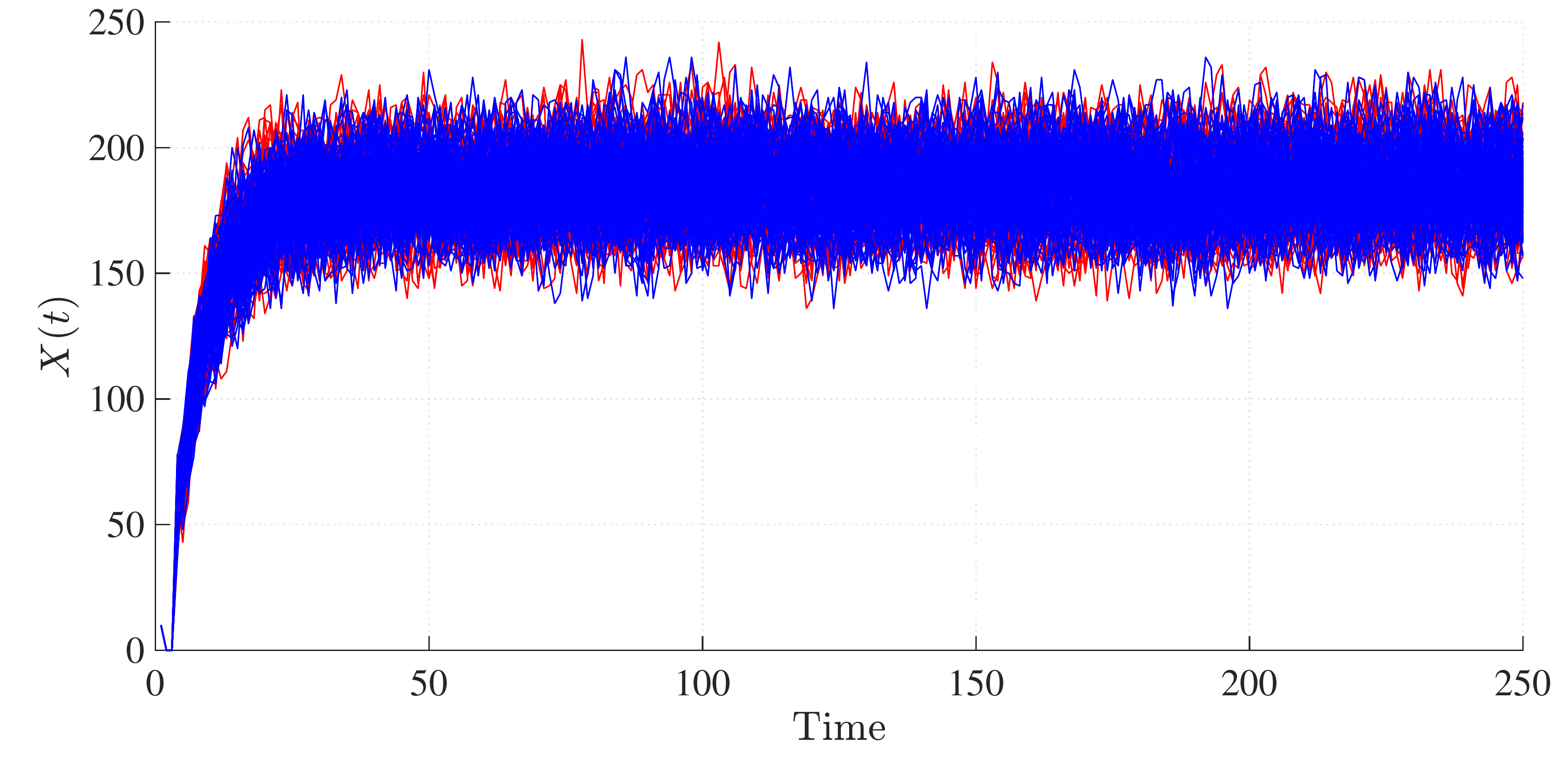}
\caption{Simulation of the Tangle. The blue realizations represent the agent based model, whereas the red realizations represent the stochastic model. The upper panel shows the leaves, $L(t)$ and the lower panel represents the free leaves, $X(t)$. Each simulation shows 500 realizations of the model with transaction times, $T_a$, generated according to a Poisson distribution with $\lambda = 60$ and $h = 3$. }
\label{Fig: Comparison}
\end{figure}

\begin{figure}
\includegraphics[width=1\columnwidth]{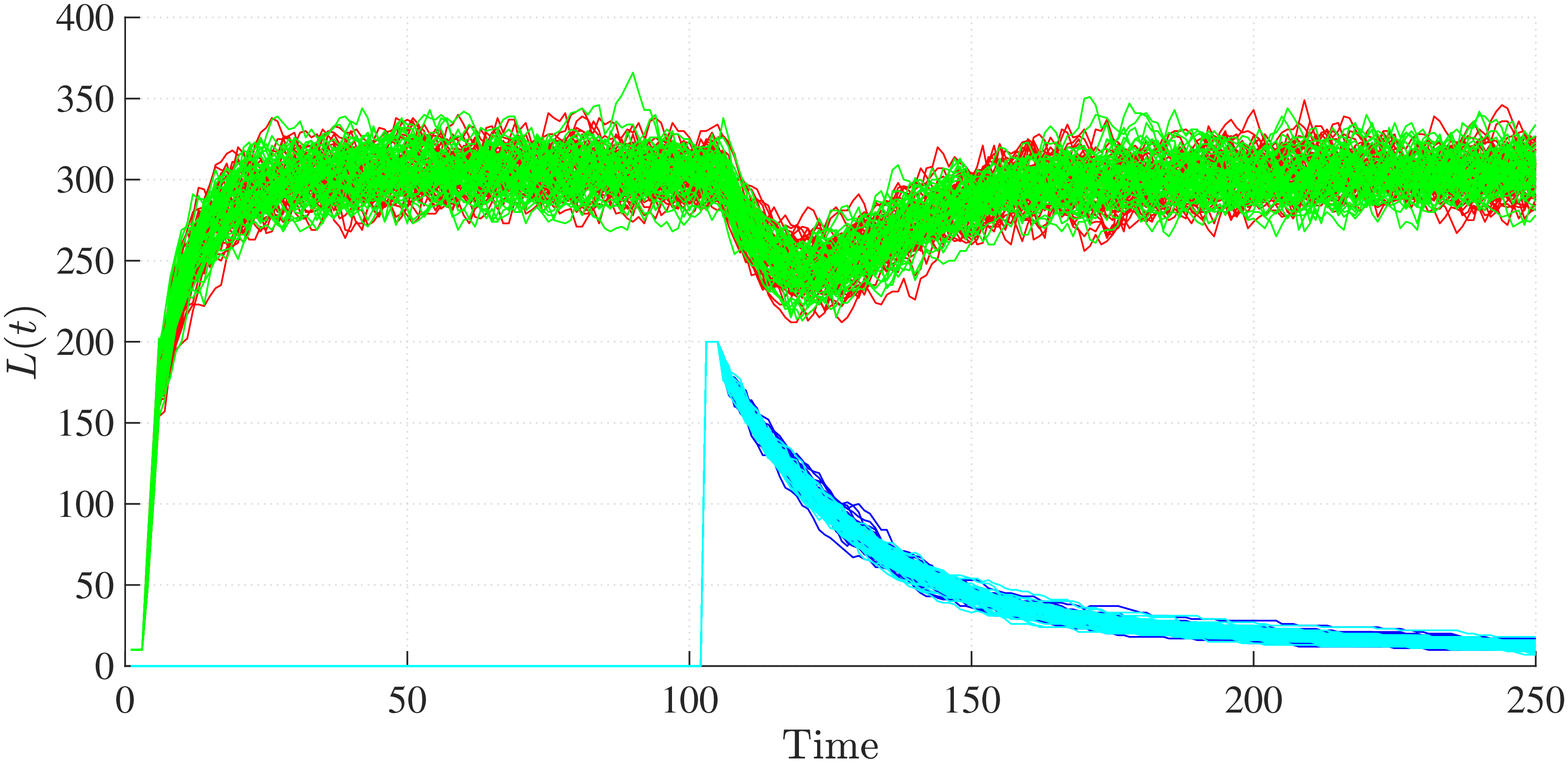}
\includegraphics[width=1\columnwidth]{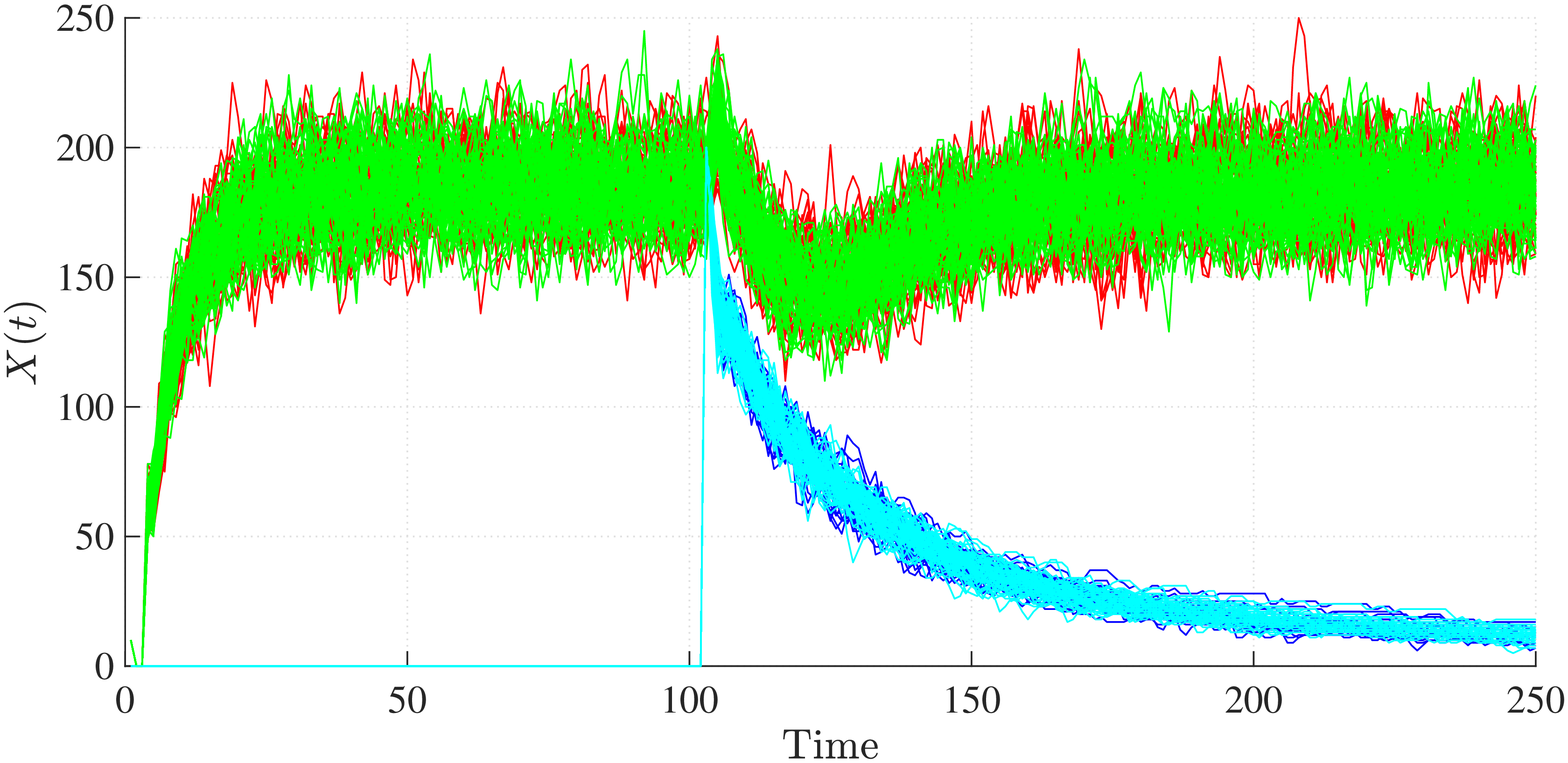}
\caption{Simulation of the Tangle with a double spending attack. The blue and red realizations represent the two types for the agent based model, whereas the cyan and green realizations represent the two types for the stochastic model. The upper panel shows the leaves, $L(t)$ and the lower panel represents the free leaves, $X(t)$. Each simulation shows 500 realizations of the model with transaction times, $T_a$, generated according to a Poisson distribution with $\lambda = 60$ and $h = 3$. At $t = 100$ the attacker issues 200 transactions trying to outpace the original Tangle.  }
\label{Fig: Comparison 2}
\end{figure}

To validate the model presented in the previous paragraph we compare its behaviour with an agent based version of the Tangle: at each time step a random number of transactions arrive, according to a Poisson distribution, and for each one of these transactions the tip selection algorithm is performed on the current tips set in order to generate graph structures equivalent to the ones presented in detail in Section \ref{Sec: DAG}. In other words, this agent based model simulates the behaviour of each transaction, therefore providing an accurate replica of the mechanism described in Section \ref{Sec: DLT structures}B.
The variables used for the comparison are the number of leaves $L(t)$ and the number of free leaves $X(t)$. To obtain these quantities, for the agent based model, it is sufficient to enumerate the number of leaves and free leaves present in the respective sets at the end of each time step. Due to the stochastic nature of the Tangle, 500 Monte Carlo simulations are performed in order to obtain statistically meaningful results. Figures \ref{Fig: Comparison} and \ref{Fig: Comparison 2} show this comparison for $d = 1$ and $d = 2$. In the second set of simulations, at $t = 100$ a user tries to perform a double spending attack on the Tangle. For both scenarios it is easy to notice, even by visual inspection, that the evolution of $L(t)$ and $X(t)$ is identical: the realizations of the agent based model are superimposed on all the realizations obtained using equations (\ref{eqn:Ni}-\ref{eqn:Li}), showing that the two systems exhibit the same dynamical behaviour. 
As a further note, notice how the attack in the second scenario fails due to the fact that not enough transactions were issued by the attacker. More details on this aspect are provided in the remainder of this Section.


\subsection{Fluid model}
In order to gain some understanding of the system (\ref{eqn:Ni}-\ref{eqn:Li}) we consider the asymptotic regime of large arrival rate,
where the time between consecutive transactions is very small. In this regime
it should be reasonable to approximate the system (\ref{eqn:Ni}-\ref{eqn:Li}) by a fluid model. 
The following derivation of the fluid model is heuristic, and it should be considered as an independent 
model for the Tangle which is motivated by the stochastic model described in the previous sections.

We introduce a scaling parameter $\lambda$ so that the arrival rate is proportional to $\lambda$,
and let $\lambda \rightarrow \infty$ to reach the fluid model. The
rescaled variables $\{\lambda^{-1} L_i(t), \lambda^{-1} X_i(t), \lambda^{-1} W_i(t)\}$ are assumed to converge
to deterministic limits as $\lambda \rightarrow \infty$, and the limits are represented
in the fluid model by real-valued functions $\{l_i(t),x_i(t),w_i(t)\}$.
The creation of new transactions in the fluid model is described by an arrival rate $a(t)$ 
which corresponds to the limit of $\lambda^{-1} N(t)$. Following this logic, the arrival rate of transactions of type $i$
in the fluid model is $a(t) p_i(t)$, where $p_i(t)$ is obtained from (\ref{prob0}) by rescaling the variables:
\be\label{def:pi}
p_i(t) = \frac{l_i(t)^{2}}{\sum_{j=1}^d l_j(t)^2}
\ee
(we assume the variables in the fluid model are continuous and thus $l_i(t - 0) = l_i(t)$ etc).
Furthermore the variable
$U(T_a)$ is replaced by its time average (over a short time interval), which
by the law of large numbers  is equivalent to the ensemble average (\ref{av-Ui}).
By rescaling variables and letting $\lambda \rightarrow \infty$, this expected value converges to
$2 x_i(t)/l_i(t)$.
Referring to the equation (\ref{eqn:Xi}) the change of $X_i$ over a small time increment $\delta$
can be approximated as
\bee
&& \hskip-0.4in X_i(t + \delta) - X_i(t)  \\
&=& N_i(t -h + \delta) - N_i(t - h) -
\sum_{a : t< T_a \le t+\delta, \, \tau_a=i} U(T_a) \\
& \simeq & \delta \, \lambda\, a(t - h) \, p_i(t - h) - \delta \, \lambda\, a(t) \, p_i(t) \, \frac{2 x_i(t)}{l_i(t)} \\
& = & \delta \, \lambda \left[a(t - h) \, p_i(t - h) - a(t) \, \frac{2 x_i(t) l_i(t)}{\sum_{j=1}^d l_j(t)^2} \right]
\eee
Applying similar reasoning to the other equations we get
the following set of delay differential equations (DDE) for the fluid model:
\be
\frac{d x_i}{d t}(t) &=& a(t - h) \, p_i(t-h) -  a(t) \, u_i(t) \label{DDEx} \\
\frac{d l_i}{d t}(t) &=& a(t - h) \, p_i(t-h) -  a(t - h) \, u_i(t-h) \label{DDEl} \\
w_i(t) &=&  l_i(t) - x_i(t) = \int_{t-h}^t a(s) \, u_i(s) d s  \label{DDEw}
\ee
where
\be\label{DDE2}
u_i(t) =   \frac{2 \, x_i(t) l_i(t)}{\sum_{j=1}^d l_j(t)^2}
\ee
The solution of these equations $\{x_i(t), l_i(t), w_i(t)\}$ can be interpreted as a fluid model which
describes the dynamics of the Tangle with very high arrival rate, using the random tip selection
algorithm. Note that
the DDE system (\ref{DDEx} - \ref{DDEw}) must be supplemented by initial conditions in the interval $[0,h]$.

\subsection{Stability of conflicts in the fluid model}
In this section we assume that the arrival rate is constant,
and we consider the existence and local stability of time-independent solutions of the fluid model.
Local stability is analyzed using a linearization of the DDE.
Given a time-independent solution $\{l_i^{(0)},x_i^{(0)},w_i^{(0)}\}$ of the
system (\ref{DDEx} - \ref{DDEw}), the linearized model is constructed by letting
\be
l_i(t) &=& l_i^{(0)} + \theta_i(t), \nonumber \\ 
x_i(t) &=& x_i^{(0)} + \xi_i(t), \nonumber \\ 
w_i(t) &=& w_i^{(0)} + \eta_i(t)
\ee
and keeping only those terms in the equations which are linear in $\theta_i, \xi_i, \eta_i$.
The resulting linear system of delay differential equations is denoted ${\cal L}^{(0)}(\theta_i, \xi_i, \eta_i)=0$.
The solution $\{l_i^{(0)},x_i^{(0)},w_i^{(0)}\}$ is locally stable if 
\be
\max_{i} \{|\theta_i(t)|, |\xi_i(t)|, |\eta_i(t)| \} \rightarrow 0 \quad \text{as $t \rightarrow \infty$}
\ee
for all solutions of the linear system. The solution is locally unstable if there is some solution
satisfying ${\cal L}^{(0)}(\theta_i, \xi_i, \eta_i)=0$ such that
\be
\max_{i} \{|\theta_i(t)|, |\xi_i(t)|, |\eta_i(t)| \} \rightarrow \infty \quad \text{as $t \rightarrow \infty$}
\ee
The spectrum of the linear system ${\cal L}^{(0)}(\theta_i, \xi_i, \eta_i)=0$ is the set of complex values
$z$ for which the system has a nonzero solution of the form
\be
(\theta_i(t), \xi_i(t), \eta_i(t)) = e^{z t} \, (\theta_i, \xi_i, \eta_i)
\ee
for some constants $(\theta_i, \xi_i, \eta_i)$. The solution is stable if the spectrum is contained in the open
left half of the complex plane, and is unstable if there is some element of the spectrum in the open right half of the
complex plane \cite{Driver}.

\begin{theorem}\label{thm1}
Consider the system  (\ref{DDEx} - \ref{DDE2}) with $d \ge 1$ types, and with arrival rate  $a(t)=1$.
For each non-empty subset $S \subset \{1,\dots,d\}$ with $|S| = k$ there is a time independent solution 
\be\label{eqn.thm1}
l_i^{(0)} = 2 \, x_i^{(0)}  = 2 \, w_i^{(0)} = \begin{cases} \frac{2 \, h}{k} & i \in S \\
0 &  i \notin S \end{cases}
\ee
For all $k = |S| > 1$ the time independent solution is locally unstable. For $k =1$ the time independent solution is locally stable.
\end{theorem}

\medskip
\noindent{\em Remark:} 
there are $2^d -1$ possible static solutions, corresponding to all possible non-empty subsets 
of $\{1,\dots,d\}$. In each static solution the total (rescaled) number of tips is $2 h$, and this total is shared equally
between the nonzero components. So the static solutions describe Tangles with equally sized conflicting sets of tips.
The linearized model describes the growth and decay of solutions which are close to the 
solutions (\ref{eqn.thm1}). So Theorem \ref{thm1} shows that all time-independent solutions are unstable except for the case where there is just one type.
This is consistent with the simulations of the full model shown in Figure \ref{Fig: Comparison 2}.

\medskip
\noindent{\em Proof:}
we look for steady state solutions of the system
(\ref{DDEx} - \ref{DDEw}):
\be
x_i(t) = x_i, \quad l_i(t) = l_i
\ee
This immediately leads to
\be
\frac{l_i^2}{\sum_j l_j^2} = 2 x_i \, \frac{l_i}{\sum_j l_j^2}
\ee
and hence
\be
l_i^2 = 2 x_i l_i
\ee
Thus either $l_i=0$ or $l_i = 2 x_i$. Similarly the relation
\be
l_i - x_i = \int_{t-h}^t u_i \, d s = u_i \, h
\ee
leads to either $x_i=0$ or
\be
\sum_j l_j^2 = 2 \, h \, l_i
\ee
Thus all nonzero components $\{l_i\}$ must be equal, and $\sum_j l_j = 2 \, h$.
Let $k\in \{1,\dots,d\}$ be the number of nonzero components of the static solution, then
these nonzero components are
\be
x_i  = \frac{h}{k}, \quad w_i = \frac{h}{k}, \quad
l_i = \frac{2 \, h}{k}
\ee

\medskip
To investigate local stability we consider small time dependent perturbations of the static solutions.
If $k < d$ there are types with $l_i^{(0)} = 0$, and these do not contribute to the linearized
equations. Thus without loss of generality we can suppose that $k=d$, so the static solution is $l_1,\dots,l_d = 2 h/d$ and
$x_1,\dots,x_d = h/d$. Consider time dependent
solutions of the form
\be
x_i(t) = \frac{h}{d} + \xi_i(t), \quad l_i(t) = \frac{2 \, h}{d} + \theta_i(t)
\ee
where $\xi_i, \theta_i$ are small. Then to linear order 
\be
u_i(t) &=& \frac{1}{d} + \frac{1}{h} \, \xi_i(t) + \frac{1}{2 \, h} \, \theta_i(t) - \frac{1}{d h} \, \sum_j \theta_j(t) \\
p_i(t) &=& \frac{1}{d} + \frac{1}{h} \, \theta_i(t) - \frac{1}{d h} \, \sum_j \theta_j(t)
\ee
and the DDE equations to linear order are
\be\label{DDE-lin}
\frac{ d \xi_i}{d t} &=& \frac{1}{h} \, \bigg[ \theta_i(t-h) - \frac{1}{d} \sum_j \theta_j(t-h) \nonumber \\
&& \hskip0.3in   - \xi_i(t) - \frac{1}{2} \theta_i(t) + \frac{1}{d} \sum_j \theta_j(t) \bigg] \nonumber \\
\frac{ d \theta_i}{d t} &=& \frac{1}{h} \, \left[ \frac{1}{2} \, \theta_i(t-h) -  \xi_i(t-h)  \right]
\ee
In order to investigate local stability, we consider a perturbation of the form
\be\label{loc-stab2}
\xi_i(t) = \xi_i e^{zt}, \quad \theta_i(t) = \theta_i e^{zt}
\ee
where $\xi_i, \theta_i$ are constant, and $z$ is a complex parameter. Substituting (\ref{loc-stab2}) into (\ref{DDE-lin}) leads to the equations
\be\label{DDE-lin-2}
(1 + h z) \, \xi_i &=& - \frac{1}{2} \theta_i + \frac{1}{d} \sum_j \theta_j + \left(\theta_i - \frac{1}{d} \sum_j \theta_j \right) e^{- z h} \nonumber \\
h z \, \theta_i &=& \left( \frac{1}{2} \theta_i - \xi_i \right) e^{- z h}
\ee
For $d > 1$ we now display a class of solutions for which $z$ is real and positive,
thus implying local instability.
To this end consider the equation
\be\label{loc-stab3}
1 + \frac{1}{2} x - e^{-x} - x e^x - x^2 e^x = 0
\ee
It can be easily checked that the left side of (\ref{loc-stab3}) is zero and increasing at $x=0$, and is negative at $x=1$.
Thus there is a positive real solution $x_0$ satisfying $0 < x_0 < 1$ (calculations show that $x_0 \simeq 0.18$). 
Define $r_0 = \frac{1}{2} - x_0 e^{x_0}$, and
let $\theta = (\theta_1,\dots,\theta_d)^T$ be any nonzero vector with
$\sum_{j=1}^d \theta_j = 0$.
Then it can be verified that there is a solution of (\ref{DDE-lin-2}) with $z = x_0 h^{-1}$ and $\xi_i = r_0 \theta_i$ for $i=1,\dots,d$.
The existence of this solution implies that the static solution is locally unstable for $d > 1$.
Conversely if $\sum_{j=1}^d \theta_j \neq 0$ then the parameter $z$ must satisfy the equation
\be\label{loc-stab4}
1 + h z = \frac{1}{2} e^{ - z h}
\ee
It can be easily checked that every solution of (\ref{loc-stab4}) lies in the open left half plane,
and thus solutions with $\sum_{j=1}^d \theta_j \neq 0$ do not lead to local instability.
This implies in particular that for $d=1$ the static solution is locally stable.

\subsection{Summary of results}
The preceding analysis strongly suggests that conflicting transactions cannot coexist on the
Tangle in the regime of large arrival rate when the random selection algorithm is used to select tips for approval. 
The fluid model does allow conflicting sub-DAGs to co-exist as long as they have exactly equal numbers of tips,
however the local instability result derived in this section implies that any small imbalance will quickly
be amplified, and we expect that this instability will eventually lead to the removal of all but one of the sub-DAGs. The time scale for this effect
is proportional to the delay parameter $h$ in the model. We conjecture that a stronger stability result holds, namely that all
solutions of the fluid model (except the ones with exactly balanced sub-DAGs) converge to the static solution with $d=1$ as $t \rightarrow \infty$.

\medskip
The fluid model is presumed
to approximate the full stochastic model for large arrival rate, and of course the stochastic model will continually exhibit
small fluctuations. Thus the locally unstable solutions of the fluid model should not appear in the stochastic model,
and this conclusion is indeed supported by our simulations of the Tangle, for example in Figure \ref{Fig: Comparison 2}. 
It is intuitively plausible that the
random selection algorithm will remove conflicts on the Tangle through this stochastic forcing mechanism, and it is
reassuring that the fluid model shows the same behavior. The paper \cite{Popov} pointed out some
potential weaknesses of the random selection algorithm which might be exploited to attack the Tangle. That paper also
proposed a different tip selection algorithm based on a random walk along the DAG, which might provide more protection
against malicious attacks. It would be interesting to analyze this random walk algorithm using the methods presented here.

\section{DLT for Social Compliance}
\label{Sec: Compliance}
We now return to the use-cases described in the introduction. We are interested in using DLT to create a privacy preserving 
mechanism to enforce social contracts in situations where machines, and other machines, or humans, must orchestrate their actions 
to ensure efficient sharing of resources. In this context the digital token is used as a bond, or digital deposit, to ensure 
that various agents comply with social contracts.  The risk of losing a token is then the mechanism that encourages agents to 
comply with these social contracts.\newline 

A natural question that arises in this context is the value associated with tokens that enforce a 
particular activity. If we set the value too low, then the risk is of no interest to an agent and the social 
contract is not enforced. If we set this value too high the activity may be suppressed entirely
for fear of losing tokens, and the societal contract also fails. Thus we wish to place a value on tokens so that 
they freely move in our system, but also so that they encourage almost 
full compliance of all agents participating in the system. 
The issue of finding this value is the subject of this present section.\newline

Before proceeding it is worth noting that the issue of compliance is sometimes not addressed in studies of algorithms to regulate, control, and optimise city infrastructures. Many publications 
addressing {\em smart city} problems assume full compliance with policies that have been engineered to optimally organise city infrastructures. For example,
papers on optimisation of traffic lights frequently do not consider the issue of compliance. Of course, humans break rules, and the effect of this 
rule-breaking often profoundly affects how cities operates. For example, readers will be certainly familiar with the effect of drivers and cyclists who break red lights or 
block traffic junctions, especially during very busy periods of road usage.\newline 

In what follows, we explore the use of our DLT strategy to regulate access to interacting junctions in a road network. A specific example could be a network of streets with traffic lights at the intersections, and a population of road users who share the space with other traffic. The compliance goal is that agents (cyclists, drivers) will be `good citizens' and will respect the instructions from the traffic lights. The control mechanism is an automatic token exchange whereby each agent arriving at a light
will deposit some number of tokens with the junction. If the agent obeys the traffic lights when crossing the intersection  then the tokens will
be returned; otherwise the tokens will be kept by the system.   {Clearly, penalties of this type are already in place in many cities using cameras. From a drivers perspective, the potential benefit of {\em losing} a deposit over, say,  camera based fines, is anonymity. Specifically, for minor indiscretions, the miscreant is penalised without a record of the offence ever being recorded, or made visible to a central entity. Of course, when all tokens expire from the wallet, this information is communicated to a municipality and the offender is no longer allowed to drive. To realise such a system we assume that car drivers have a unique digital wallet, and that tokens are exchanged with digital wallets in the fixed infrastructure in the usual manner}\protect\footnote{  {Note that in this realisation, reasonable transaction delays can be tolerated in the token exchange mechanism.}}.  The number of tokens required at the intersection can be adjusted in real time in order
to control the level of compliance. {  To give a practical example of how these tokens could work, they could be issued by the city council and sold for a fixed amount of money to every driver or they could work in a similar way to the driving license points that are used in the states of the European Union (therefore, your license is valid as long as you have a positive amount of points)}. The model assumes that each agent will  choose whether to comply with the
local traffic light based on both the cost of non-compliance (leading to lost tokens) as well as their observation of the 
behavior of other agents (the assumption being that agents are influenced by how others behave). Note in this model, we could assume, for example, that 
each road user may only use a road if he/she has a positive token balance. Once a balance goes to zero, the user is no longer allowed to travel. \newline

\begin{figure}
\includegraphics[width=1\columnwidth]{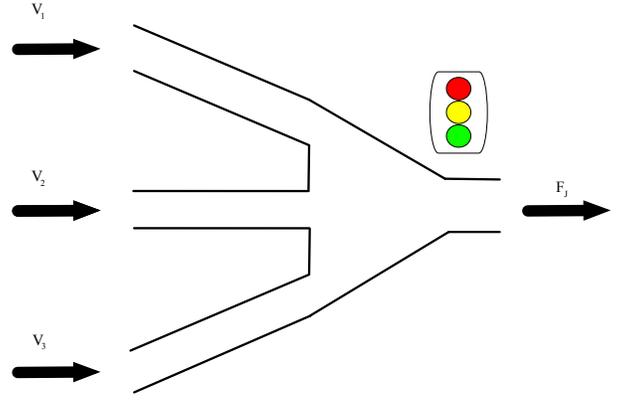}

\caption{Visual illustration of the junction example.}
\label{Fig: Junction}
\end{figure}

\begin{figure}
\includegraphics[width=1\columnwidth]{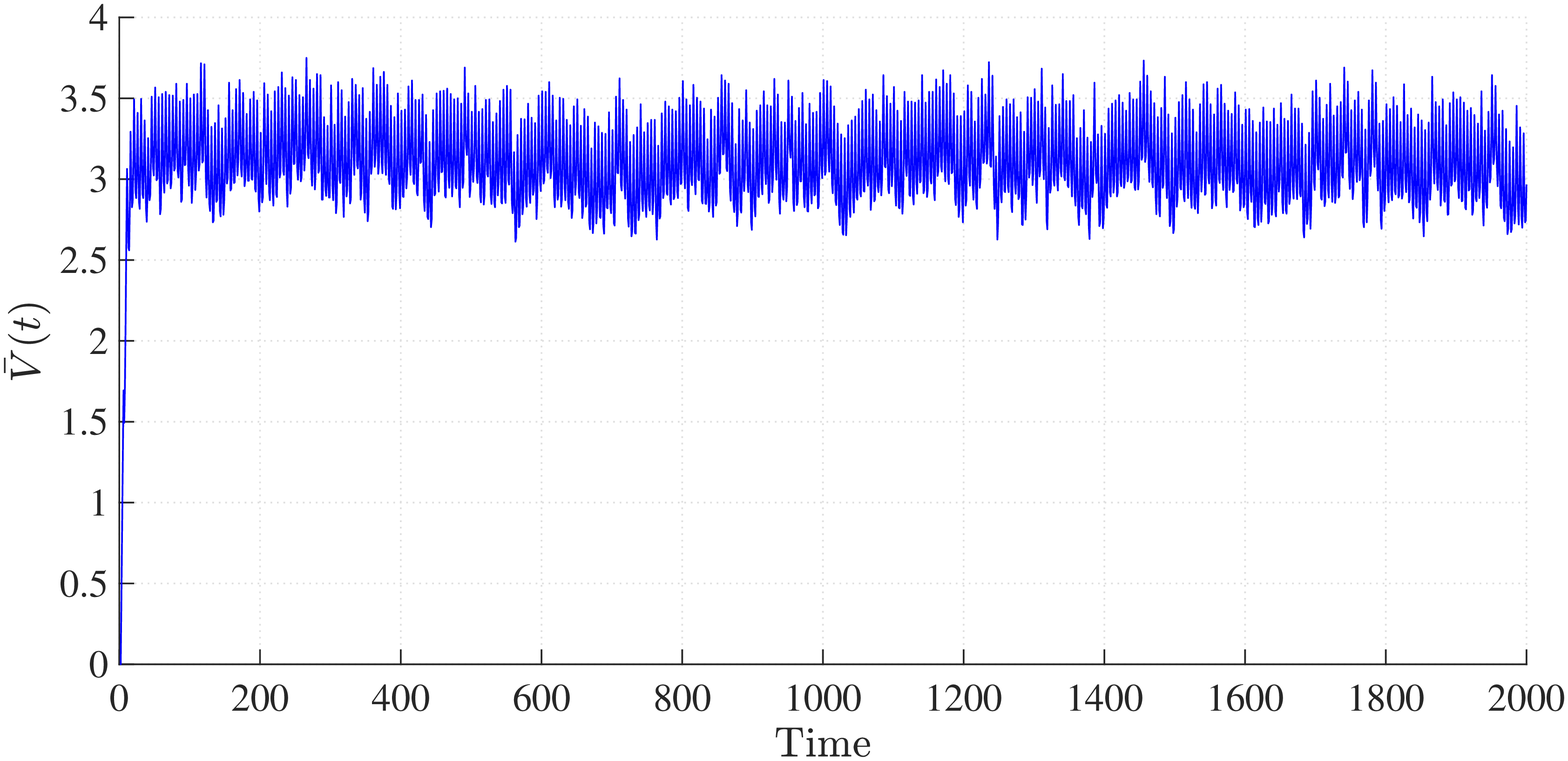}
\includegraphics[width=1\columnwidth]{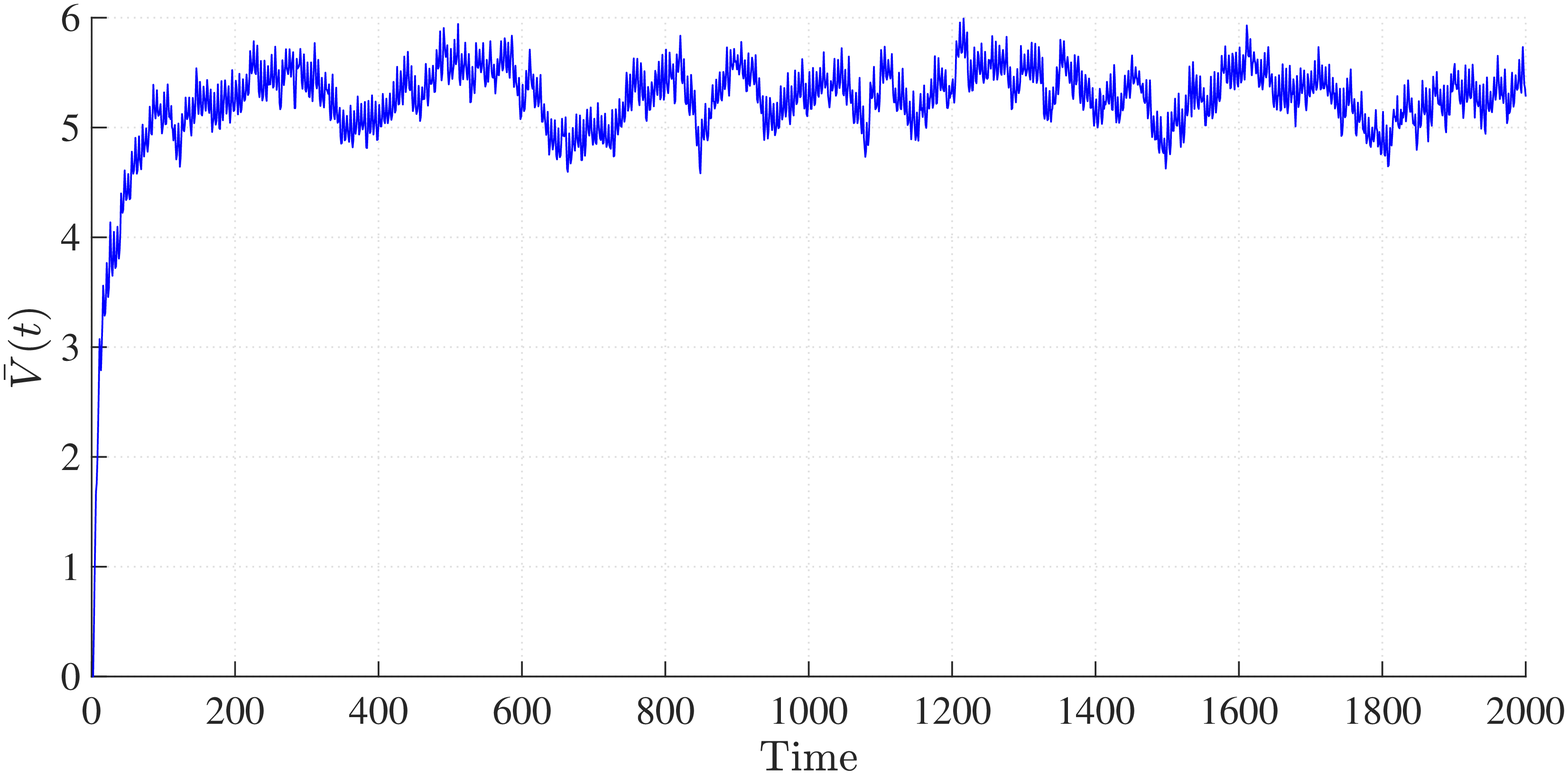}
\includegraphics[width=1\columnwidth]{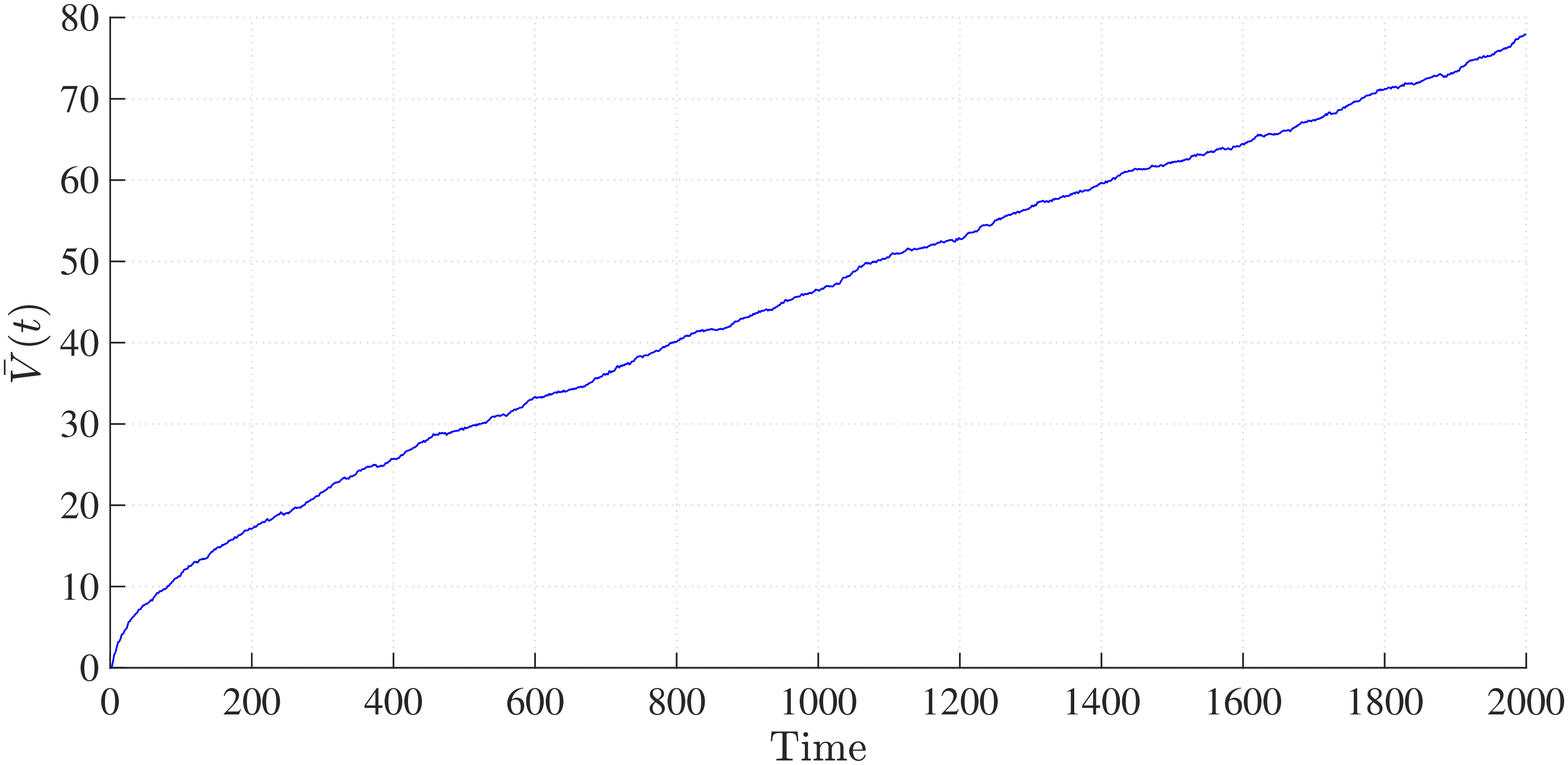}
\includegraphics[width=1\columnwidth]{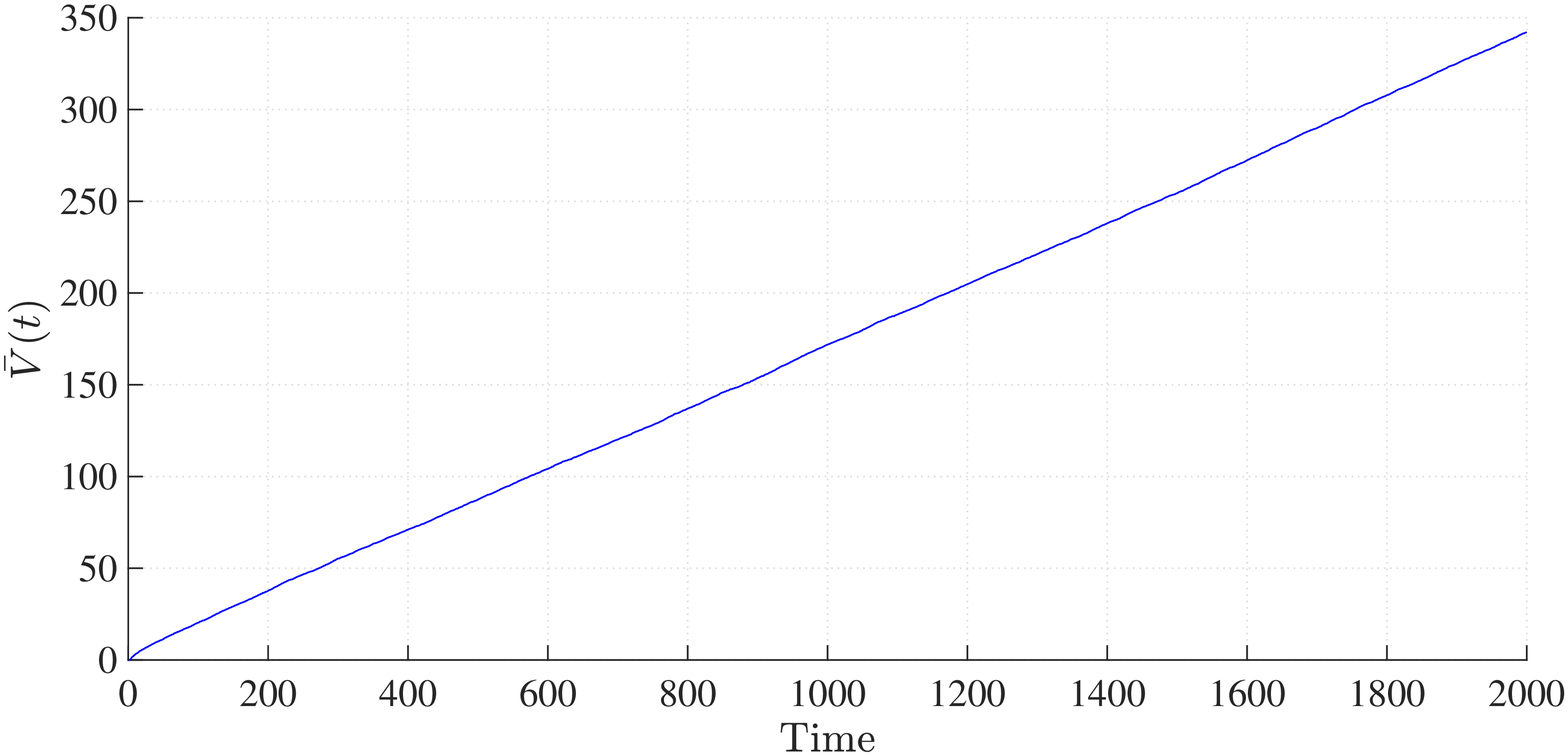}
\caption{Each image shows $\bar{V}(t)$, averaged across 100 Monte Carlo simulations. The compliance level for each simulation goes from 1, (upper panel) to 0.7 (lower panel). As the Compliance level decreases the performance of the system degrades until it becomes unstable.}
\label{Fig: Compliance increases}
\end{figure}

\begin{figure}
\includegraphics[width=1\columnwidth]{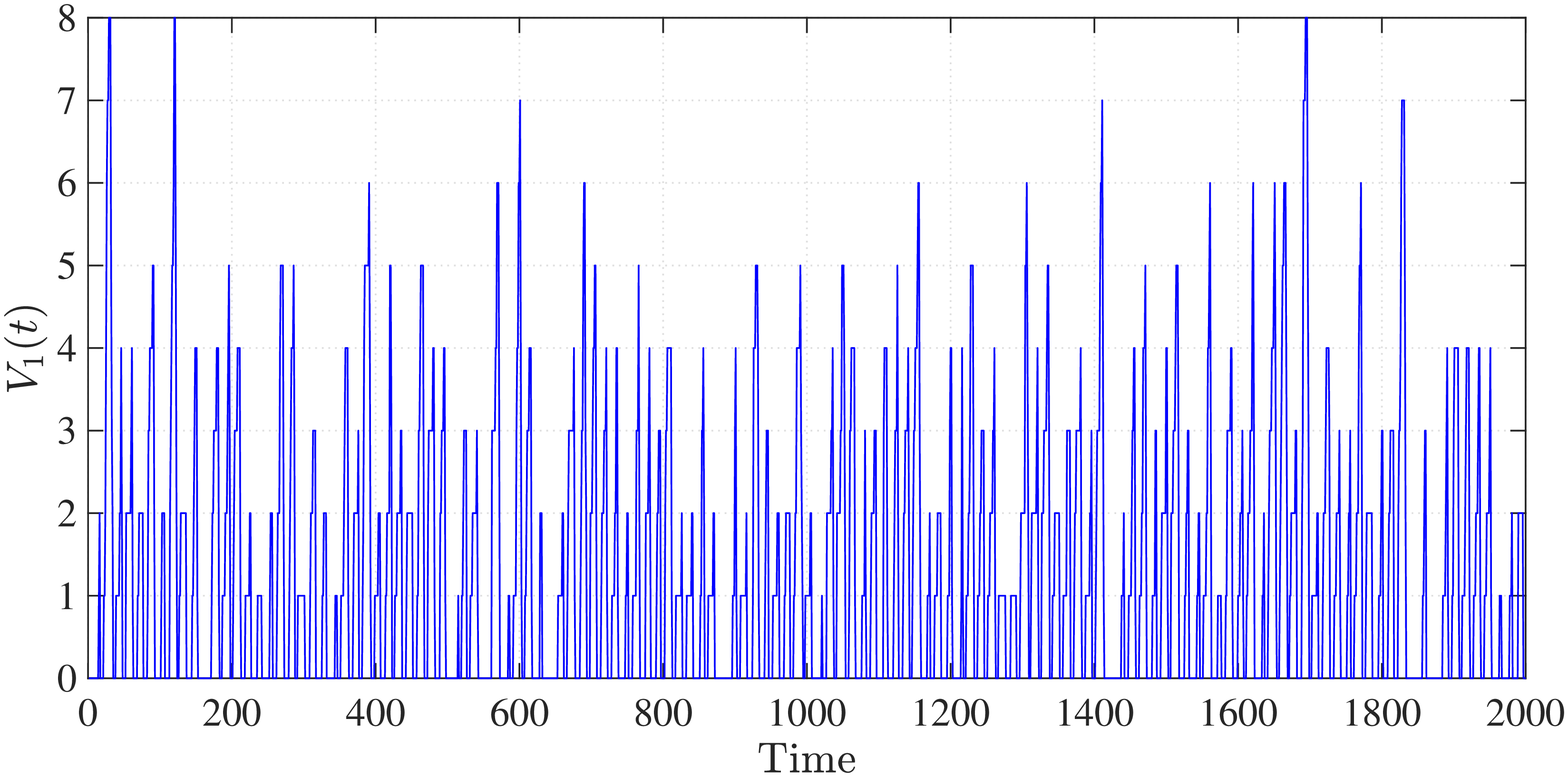}
\includegraphics[width=1\columnwidth]{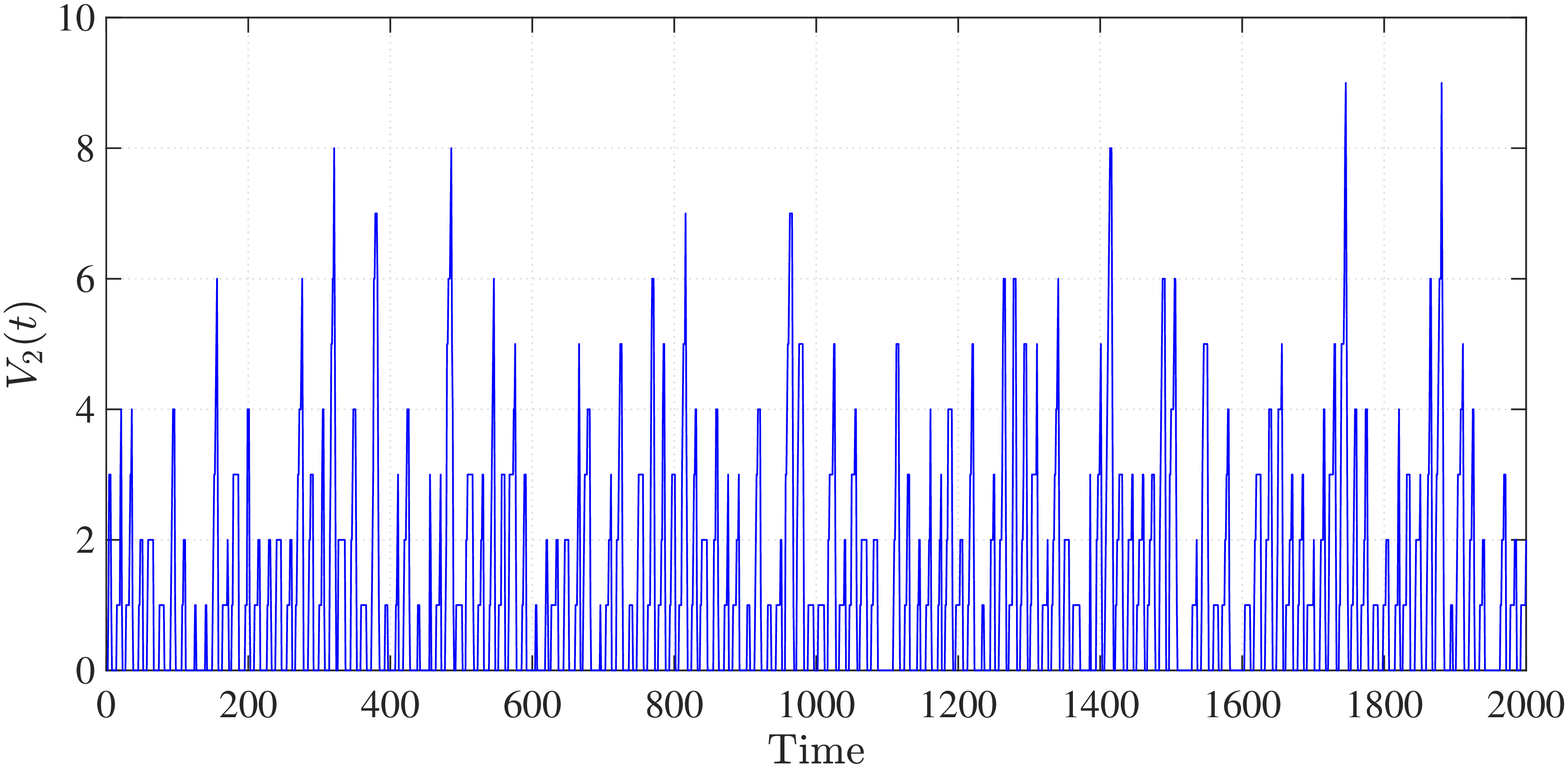}
\includegraphics[width=1\columnwidth]{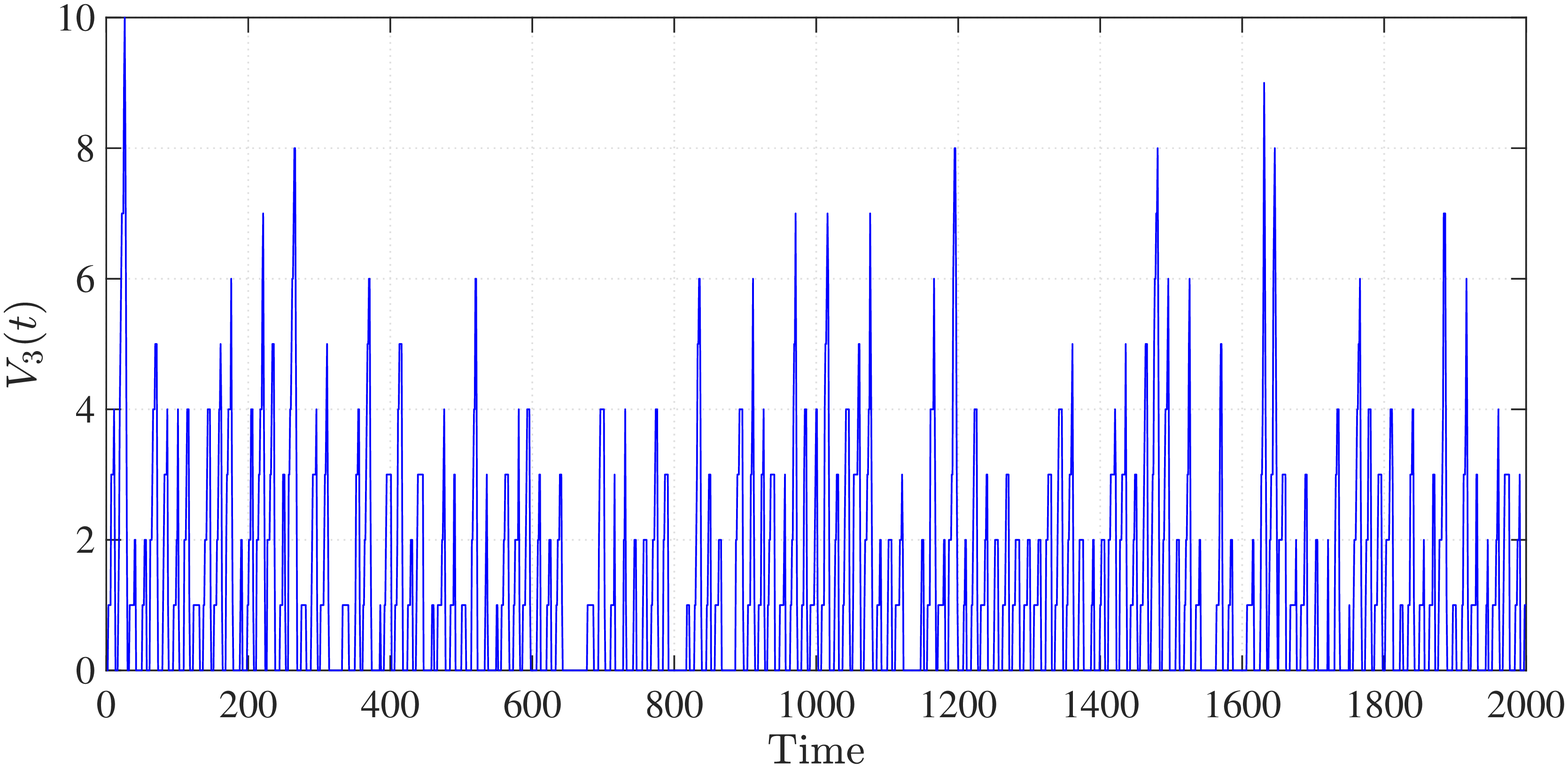}
\caption{Single realization of $V_1$, $V_2$ and $V_3$ for the junction system. }
\label{Fig: Traffic }
\end{figure}

To show why this is effectively an interesting area to focus on, in what follows we present the effects on the traffic of a single junction as the compliance rate, $Q(t)$ decreases.
Figure \ref{Fig: Junction} shows a single junction where a traffic light coordinates the traffic flow of three roads: every $T_s$ units of time the traffic light switches and allows cars from a different road to pass. The amount of time for a vehicle to cross the junction is referred to as $T_J$. The number of cars that crosses the junction, per time unit, is fixed and is referred to as $F_J$. At the same time, each time unit, a certain number of cars, $A_r$  is randomly distributed among each road, according to a uniform distribution, increasing the length of each corresponding queue. In what follows we refer to the lengths of the three queues as $V_1$, $V_2$, $V_3$ and the average queue length as $\bar{V}$, whose magnitude will be used in what follows as an estimate of the performance of the junction (i.e., the larger the average queue, the worse the junction performs). Furthermore, there is a probability, namely $1-Q(t)$, that at time $t$ a car from one of the queues with a red traffic light will go through. In this scenario, to take into account that cars are likely to slow down during this event, the crossing time increases by a fixed quantity, $\tau_d$ . To appreciate the effects of the compliance rate $Q(t)$ consider a junction with $T_s = 10$, $T_J = \tau_d = 1$, $F_J = 3$ and  $A_r $ as a random number generated  by a Poisson distribution of mean $\lambda = 1$. Figure \ref{Fig: Compliance increases} shows $\bar{V}$, averaged across 100 Monte Carlo simulations as the compliance level decreases from 1 to 0.8 (by 0.1 ticks). Note that while in the case for $Q = 1$, the junction works smoothly (Figure \ref{Fig: Traffic } shows the individual behaviour of $V_1$, $V_2$ and $V_3$ in a single realization of the process), even a small level of non compliance is enough to compromise the system performance and the delays caused by larger levels of non compliance are enough to make the system unstable. These results lead us to believe that the derivation of a control mechanism based on pricing tokens to avoid high level of non compliance, can be an effective way of improving the quality of services in cities.

The control model for this system will be described by a series of differential equations for the cost of tokens (the cost of non-compliance)
based on how the current compliance levels compare to some target levels.
Sufficient conditions are established for the existence of a locally stable solution
which achieves the target compliance goals.

\subsection{The network model}
\label{Sec: Network model}
Consider a system of $n$ physically separated activity centers (for example road junctions), and a population of agents moving between these activities.
The cost of participation at activity $i$ at time $t$ is $C_i(t)$, and the probability of compliance at
activity $i$ at time $t$ is $Q_i(t)$, with $0 \le Q_i(t) \le 1$. The time averaged compliance level $\overline{Q_i}(t)$ is defined as
\be\label{Q-bar}
\overline{Q_i}(t) = \frac{1}{w} \, \int_{t - w}^t Q_i(s)  \, d s
\ee
where $w > 0$ is a fixed window size for the average. 

\medskip
For each pair of activities $(i,j)$ the time lag $\tau_{i \rightarrow j}$ is the minimum time needed for an agent to move from
$i$ to $j$ (where $\tau_{i \rightarrow i}=0$ by definition). The compliance level at activity $i$ is assumed to depend on the cost $C_i$
as well as the time averaged compliance levels for other activities, delayed by the corresponding time lags:
for $i=1,\dots, n$
\be\label{Q-f}
Q_i(t) = f_i\left(\overline{Q_1}(t - \tau_{1 \rightarrow i}), \dots, \overline{Q_n}(t - \tau_{n \rightarrow i}), \, C_i(t)\right)
\ee
The function $f_i$ is unknowable, but for the purposes of this analysis some basic properties can be assumed: 
in particular it is assumed that $f_i$ is differentiable, and that
for all $i=1,\dots,n$ the function $f_i$ is monotone increasing in its last argument, so that
\be\label{f-mono}
\frac{\partial f_i}{\partial C_i} > 0 \quad \text{for all $i=1,\dots,n$}
\ee
Also we define the
{\em feasibility region} ${\cal F}$ to be the set of compliance levels $(Q_1,\dots,Q_n)$ which can be achieved as time-independent
solutions of (\ref{Q-f}) using some valid choice of costs $(C_1,\dots,C_n)$:
\be\label{def-feas}
{\cal F} \hskip -0.1in&=& \hskip -0.1in \{(Q_1,\dots,Q_n) \,:\, \exists \, (C_1,\dots,C_n) \in [0,\infty)^n \nonumber \\
&& \hskip0.1in   \text{s.t.} \, Q_i = f_i(Q_1,\dots,Q_n,C_i), \,\, i=1,\dots,n \}
\ee
The assumption (\ref{f-mono}) implies that for each $(Q_1,\dots,Q_n) \in {\cal F}$ there is a unique set of costs
$(C_1,\dots,C_n)$ such that the equations $Q_i = f_i(Q_1,\dots,Q_n,C_i)$ are satisfied for each $i=1,\dots,n$.

\medskip
For each activity $i$ there is a fixed target compliance level $Q_i^{(T)}$, and the control equation for the cost $C_i$ is
assumed to have the form
\be\label{C-g}
\frac{d C_i}{ d t} = g_i\left(Q_i^{(T)} - Q_i(t)\right), \quad i=1,\dots,n
\ee
The function $g_i$ is a design parameter, and in particular is chosen so that
$g_i(0)=0$, $g_i$ is differentiable and $g_i'(0) > 0$.

\medskip
To summarize, the components of the model are
\bee
C_i(t) &=& \text{cost of token to start activity $i$ at time $t$} \\
Q_i(t) & = & \text{compliance level for activity $i$ at time $t$} \\
&& \text{[$0 \le Q \le 1$, where $Q=1$ is total compliance]} \\
\overline{Q_i}(t) &=& \text{time averaged history of compliance} \\
w &=& \text{window size for time average} \\
{\cal F} &=& \text{feasibility region for compliance levels} \\
Q_i^{(T)} &=& \text{target compliance level for activity $i$ } \\
f_i( \cdots) &=& \text{compliance level function} \\
g_i( \cdot) &=& \text{cost function} \\
\tau_{i \rightarrow j} &=& \text{time lag between activities $i$ and $j$} 
\eee

{  \emph{Remark:} We want to stress the fact that while in this model the amount of deposit $C(t)$ is the same for every user, it is easy to imagine extending equations (\ref{C-g}), by taking into account variables that we neglected in our current analysis (e.g., individual wealth, amount of times the particular resource is used, amount of previous transgressions, etc.).}
\subsection{Static solution}
For a time-independent or static solution the cost $C_i$ must be constant for each $i=1,\dots,n$.
From (\ref{C-g}) and the assumptions on the functions $g_i$ it follows that the static solution must satisfy
\be
Q_i(t) = Q_i^{(T)} \quad \mbox{for all $t$, for all $i=1,\dots,n$}
\ee
These conditions are possible only if the target compliance levels are in the feasibility region, that is if
$(Q_1^{(T)},\dots,Q_n^{(T)}) \in {\cal F}$. In this case there are unique costs
$(C_1^{(T)},\dots,C_n^{(T)})$ such that 
\be\label{static1}
Q_i^{(T)} =  f_i\left(Q_1^{(T)}, \dots, Q_n^{(T)}, \, C_i^{(T)}\right), \quad i=1,\dots,n
\ee
Thus the system has a unique static solution for every set of target compliance levels in the feasibility region.

\subsection{Local stability analysis}
To investigate local stability of the solution we consider perturbations of the static solution of the form
\be
C_i(t) &=& C_i^{(T)} + \alpha_i(t), \quad Q_i(t) = Q_i^{(T)} + \beta_i(t) 
\ee
where $\alpha_i, \beta_i$ are small. Also define
\be
\theta_i(t) = \frac{1}{w} \, \int_{t - w}^t \beta_i(s)  \, d s
\ee
Then keeping only terms of lowest non-trivial order, and assuming differentiability of all functions
$\{f_i, g_i\}$ gives the system of linear equations
\be\label{lin-sys1}
\frac{d \alpha_i}{d t} &=& - g_i'(0) \, \beta_i(t) \nonumber \\
\beta_i(t) &=& \sum_{j=1}^n D_{ij} \, \theta_j(t - \tau_{j \rightarrow i}) + E_i \, \alpha_i(t)
\ee
where 
\be
D_{ij} &=& \frac{\partial f_i}{\partial \overline{Q_j}} \left(Q_1^{(T)},\dots,Q_n^{(T)},C_i^{(T)}\right) \\
E_i &=& \frac{\partial f_i}{\partial C_i} \left(Q_1^{(T)},\dots,Q_n^{(T)},C_i^{(T)}\right)
\ee
In order to investigate stability of this linear system it is sufficient to consider perturbations of the form
\be
\alpha_i(t) = \alpha_i \, e^{z t} , \quad \beta_i(t) = \beta_i \, e^{z t}
\ee
where $z$ is a complex parameter.
This assumption leads to the relations
\be\label{rels1}
\theta_j(t - \tau_{j \rightarrow i}) &=& \frac{1}{w z} \, \left(1 - e^{- w z} \right) \, e^{z (t - \tau_{j \rightarrow i})} \, \beta_j \\
\alpha_i &=& - \frac{g_i'(0)}{z} \, \beta_i
\ee
Substituting (\ref{rels1}) into (\ref{lin-sys1}) gives the eigenvalue equation for each $i=1,\dots,n$:
\be\label{lin-sys2}
\sum_{j=1}^n D_{ij} \, e^{- z \tau_{j \rightarrow i}} \, \beta_j = \frac{w}{1 - e^{- w z}} \, \left(z + E_i \, g_i'(0) \right) \beta_i
\ee
The system (\ref{lin-sys1}) is locally stable if all solutions of (\ref{lin-sys2}) occur with $z$ in the open left half plane \cite{Driver}.
In order to analyze this condition, let $M(z)$ denote the matrix with entries
\be
(M(z))_{ij} = \left(z + E_i \, g_i'(0) \right)^{-1} \, D_{ij} \, e^{- z \tau_{j \rightarrow i}}
\ee
Then letting $\beta = (\beta_1,\dots,\beta_n)^T$ denote the vector of values, the system (\ref{lin-sys2}) can be written as
\be
M(z) \, \beta = \frac{w}{1 - e^{- w z}} \,  \beta
\ee
Let $\{\lambda_a(z)\}_{a=1}^n$ denote the eigenvalues of $M(z)$, then a sufficient condition for local stability
is that all values of $z$ which for some $a \in \{1,\dots,n\}$ yield a solution of the equation
\be\label{eval2}
\lambda_a(z) = \frac{w}{1 - e^{- w z}}
\ee
should lie in the open left half plane. Although this condition involves the precise details of the entries of the
matrix $M(z)$, it is possible to find a general sufficient condition. Namely, for any $z$ in the closed right half plane,
the modulus of the right side of (\ref{eval2}) is greater than or equal to $w/2$. Thus a sufficient condition is
\be\label{cond-stab-suff}
{\rm Re}(z) \ge 0 \,\, \Rightarrow \,\, | \lambda_a(z) | < \frac{w}{2} \,\, \text{for all eigenvalues $\lambda_a$}
\ee
We will next analyze how this condition applies to a simple network.

\subsection{Special case: a ring network}
We introduce a simple network in order to analyze the stability question in more detail. So consider a network
of $n$ activities arranged in a ring, and assume that the lag time 
$\tau_{i \rightarrow j} = \tau_{j \rightarrow i} = \tau$ is constant between all pairs of
nearest neighbors $i,j$. Furthermore assume that each activity is directly influenced only by its nearest neighbors,
and that there is a constant $D > 0$ such that
\be
D_{ij} = \begin{cases} D & \mbox{if $i,j$ are nearest neighbors} \\ 0 & \mbox{otherwise} \end{cases}
\ee
Finally assume also that the values $g_i'(0)$ and $E_i$ are constant, and let
\be
\delta = E_i \, g_i'(0) > 0, \quad i=1,\dots,n
\ee
Then the eigenvalues of the matrix $M(z)$ are
\be
\lambda_a(z)= \frac{D e^{- z \tau}}{z + \delta} \, 2 \cos \left(\frac{2 \pi a}{n}\right), \quad a=1,\dots,n
\ee
Letting $z = x + i y$ with $x \ge 0$, we have the bound
\be
| (\lambda_a(z)) | \le \frac{2 D}{\delta}
\ee
Therefore (\ref{cond-stab-suff}) implies that a sufficient condition for local stability is
\be\label{ring-stab-cond}
D < \frac{w \delta}{4}
\ee
The parameter $D$ records the sensitivity of the compliance functions (\ref{Q-f}) to the behavior
of their neighboring activities. So for fixed $w, \delta$ the condition (\ref{ring-stab-cond}) requires that agents are not unduly influenced
by the behavior of agents at other locations. For more complicated networks the same general result applies, namely that
weak influence between agents at different activities leads to local stability of the static solution.

\subsection{Example: single junction}

\begin{figure}
\includegraphics[width=1\columnwidth]{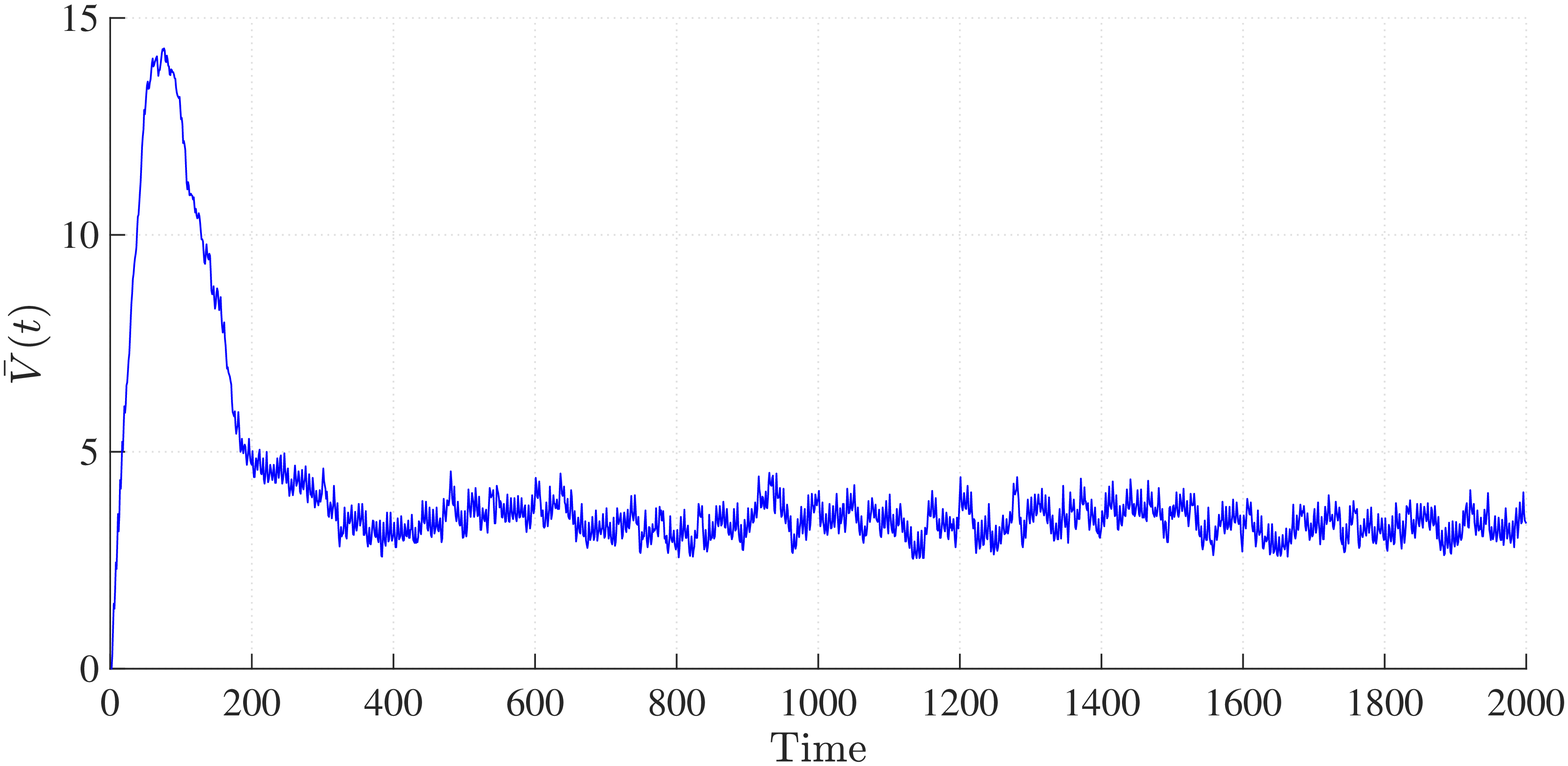}
\includegraphics[width=1\columnwidth]{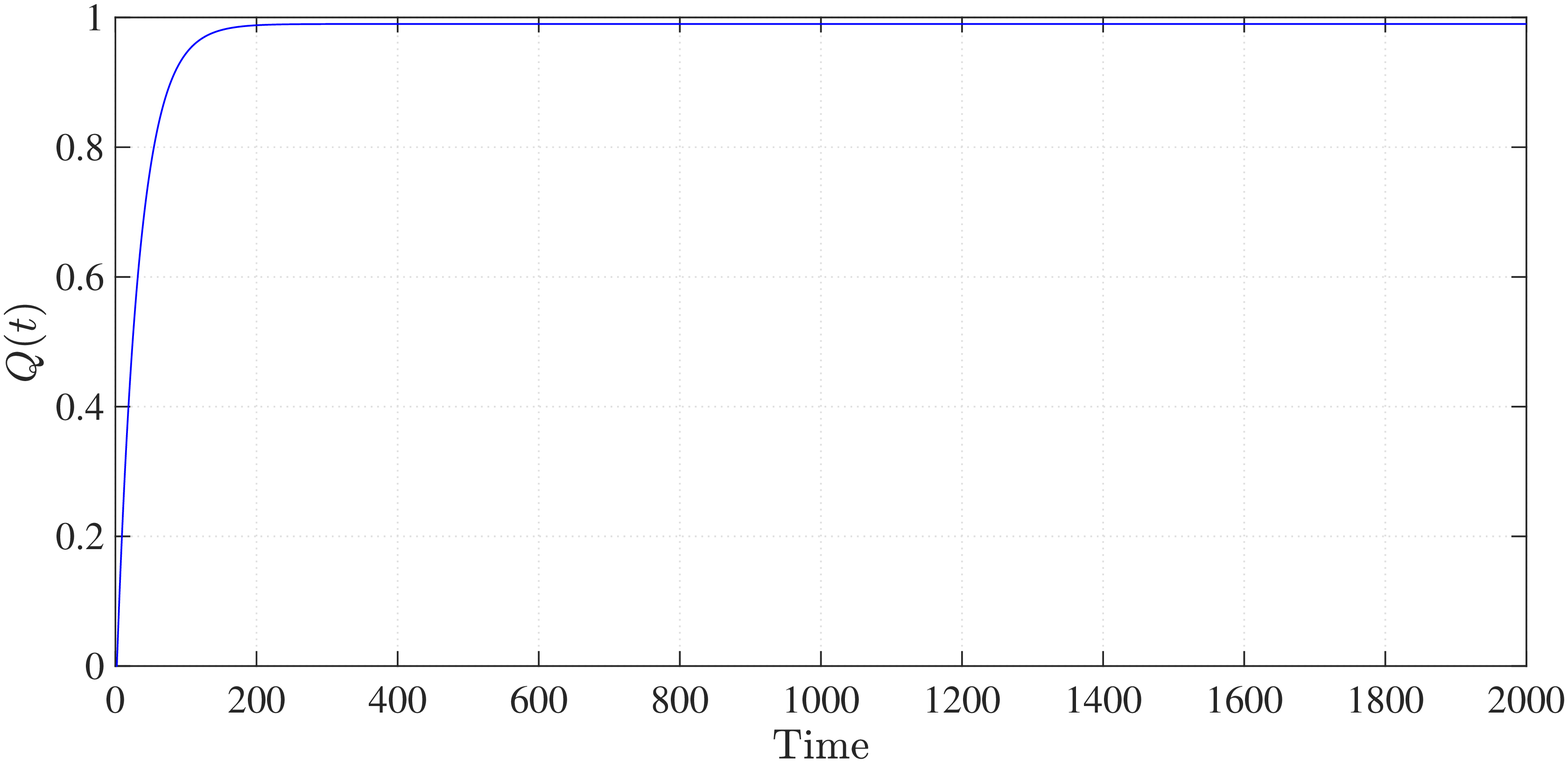}
\includegraphics[width=1\columnwidth]{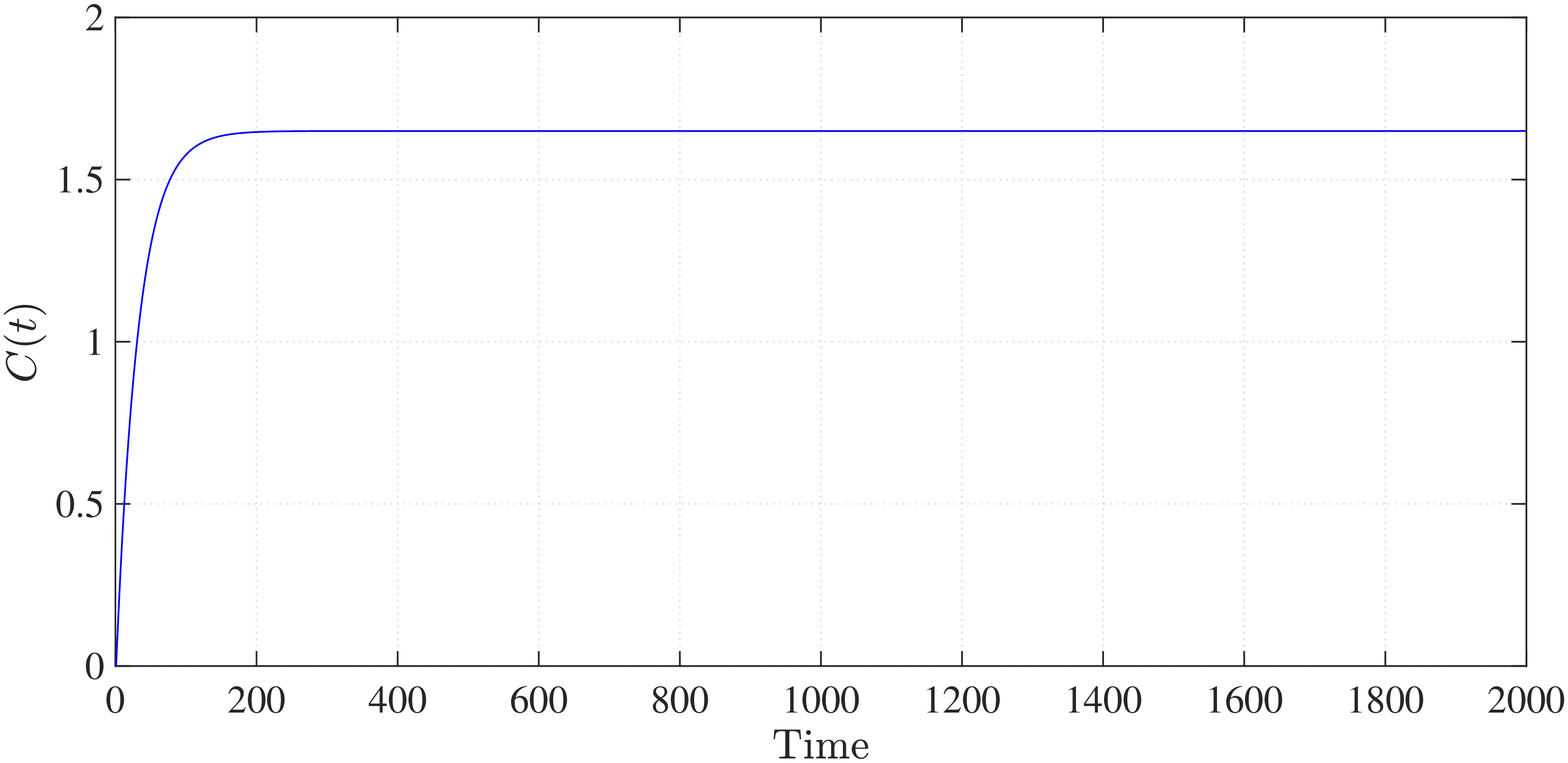}
\caption{Simulation of the junction system, with $ \beta = 0.6, \tau_C = 1, K = 0.1, \bar{V}(0) = C(0) = 0$. The upper panel shows the average queue length, $\bar{V}(t)$, the middle panel shows the compliance level $Q(t)$ and the lower panel show the cost, $C(t)$. }
\label{Fig: Control }
\end{figure}

We now return to the single junction introduced in Section \ref{Sec: Compliance}. In order to prevent the drastic decrease in performance due to the compliance level and to maintain the traffic of the junction within acceptable levels (with respect to the optimal level of compliance), we make use of the  token system described in Section \ref{Sec: Network model}: assume, for the sake of simplicity, that the dynamics of the cost, $C(t)$  are chosen as a first order system and that the compliance rate depends linearly on $C(t)$:

\be
Q(t + 1) =  \beta C(t + 1),
\label{Eq: Compliance example}
\ee
\be
C(t+1) = \tau_C C(t) + K(Q^{(T)}-Q(t)),
\label{Eq: Cost example}
\ee
where $t$ represents the time unit of the junction system, the parameter $\beta$ represents values that can be empirically extracted by data collected at the junction and $\tau_C$ and $K$ represent the design parameters for the control system. Figure \ref{Fig: Control } shows $\bar{V}(t)$ (upper panel), $Q(t)$ (middle panel) and $C(t)$ (lower panel): after an initial transient (where the oscillations are emphasized for illustrative purposes) $Q(t)$ stabilizes around the desired compliance rate (0.95, in this example). Furthermore, note that the steady state value of $C(t)$ depends on the parameters of equation (\ref{Eq: Compliance example}) (specifically on the choice of $\beta$). These values are supposed to be extracted empirically from real data (see equation (\ref{Q-f})) and therefore it needs to be stressed that the particular steady state value reached by $C(t)$ depends specifically on the arbitrary choice for this example.
 \newline

\section{Conclusions and Future Work}
\label{Sec: Conclusions}

The contributions of the paper are divided into two main parts. In the first one we present a stochastic model for one choice of
a DAG-based DLT (the Tangle), and demonstrate its validation through an extensive Monte Carlo analysis. The results of the validation show that our model matches with a high degree of accuracy the real Tangle. Then, under the assumption of a high arrival rate of transactions, we derive a deterministic version of the stochastic model and show, through a local instability theorem, that conflicting transactions can not coexist when a random tip-selection algorithm is employed. The second part presents a general framework to apply DLTs as a control mechanism for social compliance in smart city environments: first we show, through the example of a junction with a traffic light, how even small levels of non compliance can lead such a simple system to instability. This example is then generalized to a generic set of interconnected activities in a road network to which the DLT-based control mechanism is applied. Finally, we present sufficient conditions for the stability of the controlled system. { As for future lines of research, the present work can be extended in a series of directions: the current Tangle model takes into account constant delays and a random selection algorithm for tip; it would be interesting to relax these assumptions to analyse the Tangle behaviour with variable delays and different tips selection strategies. Moreover the model for social compliance can be further elaborated by taking into account different factors that concur to determine the price $C(t)$ (e.g. individual wealth, amount of times the particular resource is used, previous transgressions, feasibility of the price, etc.) and it would be interesting to study whether or not the system exhibits  a stable behaviour under these conditions.}
\section{Acknowledgements}
The work is partly supported by the Danish ForskEL programme (now EUDP) through the Energy Collective project (grant no. 2016- 1-12530) and by SFI grant 16/IA/4610.


\begin{thebibliography}{88}

\bibitem{Walport} Walport, M. G. C. S. A., ``Distributed ledger technology: Beyond blockchain", {\em UK Government Office for Science, 1 Victoria Street
London SW1H 0ET}, No. GS/16/1, 2016
\bibitem{Olnes} Olnes, S., Ubacht, J. and Janssen, M., ``Blockchain in government: Benefits and implications of distributed ledger technology for information sharing", {\em Government Information Quarterly}, Vol. 34, No. 3, pp. 355-364, 2017
\bibitem{SpringerChina} Guo, Y.  and Liang, C.,  ``Blockchain application and outlook in the banking industry", {\em Financial Innovation}, Vol. 2, No. 1, pp. 24, 2016
\bibitem{EVAnxiety}  O'Connell , J., Cardiff, B. and Shorten, R., ``dockChain: A Solution for Electric Vehicles Charge Point Anxiety", In Proceedings of {\em 21st IEEE International Conference on Intelligent Transportation Systems}, 2018
\bibitem{sharing1} Griggs, W.,  Yu, J.,  Wirth, F., Haeusler, F.,  Shorten, R.,, ``On the Design of Campus Parking Systems with QoS guarantees",  IEEE Transactions on Intelligent Transportation Systems, 17(5), pp. 1428-1439, 2016
\bibitem{sharing2} King, C., Griggs, W., Wirth, F., Quinn, K., and Shorten, R., ``Alleviating a form of electric vehicle range anxiety through On-Demand vehicle access", International Journal of Control, Vol 4, pp. 717-728, 2014.
\bibitem{sharing3} Wirth, F, Stuedli, S., Yu, J., Corless, M., Shorten, R., ``Non-homogeneous Markov Chains, unsynchronised AIMD, and Optimization", {\em arXiv preprint} arXiv:1404.5064v4, 
 submitted to Journal of the ACM, 2017
\bibitem{sharing4} Fioravanti, A. Marecek, J., Shorten, R., Souza, M. Wirth, F., ``On the Ergodic Control of Ensembles", {\em arXiv preprint} arXiv:1807.03256, submitted to Automatica, 2018
\bibitem{Popov} Popov, S., ``The Tangle-Version 1.4.2", available at \url{https://iota.org/IOTA_Whitepaper.pdf}, 2017
\bibitem{Nakamoto} Nakamoto, S., ``Bitcoin: A peer-to-peer electronic cash system", available at \url{https://bitcoin.org/bitcoin.pdf}, 2008
\bibitem{Puthal} Puthal, D., Malik, N., Mohanty, S. P., Kougianos, E., and Das, G., ``Everything You Wanted to Know About the Blockchain: Its Promise, Components, Processes, and Problems" {\em IEEE Consumer Electronics Magazine}, Vol. 7, No.4, pp. 6-14, July 2018
\bibitem{Conoscenti} Conoscenti, M., Vetro, A. and De Martin, J. C., ``Blockchain for the Internet of Things: A systematic literature review", {\em  IEEE/ACS 13th International Conference on Computer Systems and Applications},  pp. 1-6, November 2016
\bibitem{Zheng} Zheng, Z., Xie, S., Dai, H., Chen, X. and Wang, H., ``An overview of Blockchain technology: Architecture, consensus, and future trends" {\em IEEE International Congress on Big Data}, pp. 557-564, 2017
\bibitem{Tangle} Popov, S., Saa, O. and Finardi, P., ``Equilibria in the Tangle" {\em arXiv preprint} arXiv:1712.05385, 2017
\bibitem{Gatteschi} Gatteschi, V., Lamberti, F., Demartini, C., Pranteda, C. and Santamaria, V., ``To Blockchain or Not to Blockchain: That Is the Question",  {\em IEEE IT Professional}, Vol. 20, No. 2, pp. 62-74, 2018
\bibitem{Puthal2}Puthal, D., Malik, N., Mohanty, S., Kougianos, E., and Yang, C., ``The Blockchain as a decentralized security framework,? {\em IEEE Consumer Electronics Magazine}, vol. 7, No. 2, pp. 18?21, 2018.
\bibitem{Healthcare} Esposito, C., De Santis, A., Tortora, G., Chang, H. and Choo, K. K. R., ``Blockchain: A Panacea for Healthcare Cloud-Based Data Security and Privacy?"  {\em IEEE Cloud Computing}, Vol. 5, No. 1, pp. 31-37, 2018
\bibitem{PKI} Orman, H.  ``Blockchain: the Emperors New PKI?", {\em IEEE Internet Computing}, Vol. 22, No. 2, pp. 23-28, 2018

\bibitem{Novo} Novo, O., ``Blockchain Meets IoT: an Architecture for Scalable Access Management in IoT",  {\em IEEE Internet of Things Journal}, Vol. 5, No. 2, pp. 1184-1195, 2018

\bibitem{srikant} Srikant, R., ''The mathematics of Internet Congestion Control",, Birkh{\"{a}}user, 2003

\bibitem{book}  Crisostomi, E., Shorten, R., St\"{u}dli, S.,. Wirth, F., ''Electric and Plug-in Hybrid Vehicle Networks: Optimization and Control", CRC Press, 2017.

\bibitem{arieh} Schlote, A., King, C., Crisostomi, E. and Shorten, R., ''Delay-tolerant stochastic algorithms for parking space assignment", 
	{\em IEEE Transactions on Intelligent Transportation Systems}, Vol. 5, No. 15, pp. 1922-1935, 2014.

\bibitem{cdc} Fioravanti, A., Marecek, J., Shorten, R. Souza, M., Wirth, F., "On classical control and smart cities", {\em arXiv preprint} arXiv:1703.07308, 2017.

\bibitem{Phan} Phan, T., Annaswamy, A. M., Yanakiev, D. and Tseng, E., ``A model-based dynamic toll pricing strategy for controlling highway traffic" {\em IEEE American Control Conference (ACC),} pp. 6245-6252, 2016

\bibitem{Soylemezgiller} Soylemezgiller, F., Kuscu, M. and Kilinc, D., ``A traffic congestion avoidance algorithm with dynamic road pricing for smart cities. In Personal Indoor and Mobile Radio Communications ", {\em IEEE 24th Annual International Symposium on Personal, Indoor, and Mobile Radio Communications (PIMRC) }, pp. 2571-2575, 2013

\bibitem{Bui} Bui, K. T., Huynh, V. A., and Frazzoli, E., `` Dynamic traffic congestion pricing mechanism with User-Centric considerations", {\em 15th International IEEE Conference on Intelligent Transportation Systems}, pp. 147-154, 2012
\bibitem{Annaswamy}Annaswamy, A. M., Guan, Y., Tseng, H. E., Zhou, H., Phan, T. and Yanakiev, D, ``Transactive Control in Smart Cities",  {\em Proceedings of the IEEE}, Vol. 106, No. 4, pp. 518-537, 2018
\bibitem{Widergren} Widergren, S., Fuller, J., Marinovici, C. and Somani, A., ``Residential transactive control demonstration", {\em IEEE PES Innovative Smart Grid Technologies Conference (ISGT)}, pp. 1-5, 2014
\bibitem{Huang} Huang, P., Kalagnanam, J., Natarajan, R., Hammerstrom, D., Melton, R., Sharma, M. and Ambrosio, R., ``Analytics and transactive control design for the pacific northwest smart grid demonstration project"  {\em First IEEE International Conference on Smart Grid Communications (SmartGridComm),} pp. 449-454, 2010
\bibitem{Kotb} Kotb, A. O., Shen, Y. C., Zhu, X. and Huang, Y., ``iParker-A New Smart Car-Parking System Based on Dynamic Resource Allocation and Pricing",  {\em IEEE Transactions on Intelligent Transportation Systems,} Vol. 17, No. 9, pp. 2637-2647, 2016
\bibitem{Yao}Yao, Y. and Zhang, P., Transactive control of air conditioning loads for mitigating microgrid tie-line power fluctuations", {\em IEEE PES General Meeting}, pp. 1-1,2016
\bibitem{Katipamula} Katipamula, S., ``Smart buildings can help smart grid: Transactive controls"{\em IEEE PES Innovative Smart Grid Technologies (ISGT)}, pp. 1-1, 2012
\bibitem{Junjie} Junjie, H. U., Guangya, Y. A. N. G., Koen, K. O. K., Yusheng, X. U. E. and Bindner, H. W., ``Transactive control: a framework for operating power systems characterized by high penetration of distributed energy resources"  {\em Journal of Modern Power Systems and Clean Energy}, Vol. 5, No. 3, pp. 451-464, 2017
\bibitem{Li} Li, J., Lin, X., Nazarian, S. and Pedram, M., ``CTS2M: concurrent task scheduling and storage management for residential energy consumers under dynamic energy pricing"  {\em IET Cyber-Physical Systems: Theory and Applications}, Vol. 2, No. 3, pp. 111-117, 2017

\bibitem{Chekired} Chekired, D. A., Khoukhi, L. and Mouftah, H. T., ``Decentralized cloud-SDN architecture in smart grid: A dynamic pricing model"  {\em IEEE Transactions on Industrial Informatics}, Vol. 14, No. 3, pp. 1220-1231, 2018
\bibitem{Bejestani} Bejestani, A. K., Annaswamy, A. and Samad, T., ``A hierarchical transactive control architecture for renewables integration in smart grids: Analytical modeling and stability", {\em IEEE Transactions on Smart Grid}, Vol. 5, No. 4, pp. 2054-2065, 2014
\bibitem{Hao} Hao, H., Corbin, C. D., Kalsi, K. and Pratt, R. G., ``Transactive control of commercial buildings for demand response"  {\em IEEE Transactions on Power Systems}, Vol. 32, No. 1, pp. 774-783, 2017

\bibitem{Swan} Swan, M., ``Blockchain: Blueprint for a new economy"  {\em O'Reilly Media, Inc.}, 2015


\bibitem{Hash} Ramakrishna, M. V., Fu, E. and Bahcekapili, E., ``Efficient hardware hashing functions for high performance computers" {\em IEEE Transactions on Computers}, Vol. 46, No. 12, pp. 1378-1381, 1997
\bibitem{Blockchain Security 1} Karame, G. O. and Androulaki, E., ``Bitcoin and blockchain security", {\em Artech House}, 2016
\bibitem{Blockchain Security 2} Karame, G., ``On the security and scalability of bitcoin's blockchain" {\em Proceedings of the 2016 ACM SIGSAC Conference on Computer and Communications Security}, pp. 1861-1862, 2016
\bibitem{Blockchain Security 3} Li, X., Jiang, P., Chen, T., Luo, X. and Wen, Q., ``A survey on the security of Blockchain systems", {\em Elsevier, Future Generation Computer Systems,} available at \url{https://doi.org/10.1016/j.future.2017.08.020}, 2017
\bibitem{Blockchain Security 4} Gervais, A., Karame, G. O., Wüst, K., Glykantzis, V., Ritzdorf, H. and Capkun, S., ``On the security and performance of proof of work blockchains", {\em Proceedings of the 2016 ACM SIGSAC Conference on Computer and Communications Security}, pp. 3-16, 2016
\bibitem{Ouroboros} Kiayias A., Russell A., David B. and Oliynykov R, ``Ouroboros: A Provably Secure Proof-of-Stake Blockchain Protocol", {\em Advances in Cryptology - CRYPTO 2017, Lecture Notes in Computer Science}, Vol. 10401, Springer, Cham, 2017 
\bibitem{Bramas} Bramas, Q., ``The Stability and the Security of the Tangle", hal-01716111v2, 2018
\bibitem{DDE} Kuang, Y. (Ed.), ``Delay differential equations. Boston", {\em MA: Academic Press}, 1993
\bibitem{Driver} R. D. Driver, ``Ordinary and delay differential equations'', {\em Applied Mathematical Sciences} 20, New York, NY, USA:Springer-Verlag, 1977

\end{thebibliography}
\end{document}